\documentclass[aps,prl,noeprint,superscriptaddress,reprint]{revtex4-1}
\usepackage{graphicx}
\usepackage{braket}
\usepackage{amsmath}
\usepackage{amsthm}  
\usepackage[british]{babel}  
\usepackage[hidelinks]{hyperref}  
\graphicspath{{../figs/}}

\begin{document}

\title{$\mathcal{PT}$-Symmetric Topological Edge-Gain Effect}

\author{Alex Y.\ Song}
\affiliation{Department of Electrical Engineering, Stanford University, Stanford, CA 94305, USA}
\author{Xiao-Qi Sun}
\affiliation{Department of Physics, Stanford University, Stanford, CA 94305, USA}
\author{Avik Dutt}
\affiliation{Department of Electrical Engineering, Stanford University, Stanford, CA 94305, USA}
\author{Momchil Minkov}
\affiliation{Department of Electrical Engineering, Stanford University, Stanford, CA 94305, USA}
\author{Casey Wojcik}
\affiliation{Department of Electrical Engineering, Stanford University, Stanford, CA 94305, USA}
\author{Haiwen Wang}
\affiliation{Department of Applied Physics, Stanford University, Stanford, CA 94305, USA}
\author{Ian A. D. Williamson}
\affiliation{Department of Electrical Engineering, Stanford University, Stanford, CA 94305, USA}
\author{Meir Orenstein}
\affiliation{Department of Electrical Engineering, Technion - Israel Institute of Technology, Technion City, Haifa 3200003, Israel}
\author{Shanhui Fan}
\email[]{shanhui@stanford.edu}
\affiliation{Department of Electrical Engineering, Stanford University, Stanford, CA94305, USA}

\date{\today}
\let\oldDelta\Delta
\renewcommand{\Delta}{\text{\scalebox{0.75}[1.0]{$\oldDelta$}}}
\newcommand{\llangle}{\langle\!\langle}
\newcommand{\rrangle}{\rangle\!\rangle}

\begin{abstract}
    We demonstrate a non-Hermitian topological effect that is characterized by having complex eigenvalues only  in the edge states of a topological material, despite the fact that the material is completely uniform.
    Such an effect can be constructed in any topological structure formed by two gapped sub-systems, e.g., a quantum spin-Hall system, with a suitable non-Hermitian coupling between the spins.
    The resulting complex-eigenvalued edge state is robust against defects due to the topological protection.
    In photonics, such an effect can be used for the implementation of topological lasers, in which a uniform pumping provides gain only in the edge lasing state. 
    Furthermore, such a topological lasing model  is reciprocal and is thus compatible with  standard photonic platforms.
    
\end{abstract}


\maketitle

The hallmark of an electric topological insulator is the striking contrast between its edge and its bulk characteristics
\cite{Qi2011,Hasan2010,Maciejko2011}.
There, the bulk is in a gapped insulating phase with a band structure having a non-trivial topology. As a result of such a non-trivial topology, the system supports gap-spanning edge states that carry charge or spin currents, and hence the edge is metallic \cite{Kane2005a,Kane2005,BHZ2006,Konig2007,Haldane1988,Thouless1982,Hatsugai1993}. 
Similar contrast exists for many photonic topological systems as well \cite{Khanikaev2017,RevModPhys.91.015006,Lu2016,Lu2014,Liang2013a,Hafezi2011,Hafezi2013,Mittal_Hafezi_2016,Yang2019,Lin2018,Nalitov2015,Kort-Kamp2017,Rechtsman2013,Khanikaev2013,Bliokh2015,Onoda2004,Wang2009,Barik2018,Haldane2008}. 

Most of the initial works on topological electronic or photonic systems assume a Hermitian Hamiltonian. On the other hand, there has been significant recent interest in exploring the topological edge states of non-Hermitian systems where gain and/or loss are present, due to their potential applications in wave control, information processing, and robust lasing \cite{Liu2019,Kawabata2018, Leykam2017, Zeuner2015, Weimann2017,Liang2013, Kunst2018, Shen2018, Esaki2011, Malzard2015, Dangel2018, CLEO_Hughes_Rechtsman_8426563,Kawabata2019a,Hirsbrunner2019p, Kremer2019,Gong2018, Yao2018, HassaniGangaraj2018, Yuce2018,Harari2018, Secli2019p, Takata2018,Bahari2017,Ni2018,Hou2019,Yuce2018a,Hu2011}. 
For these applications, it would be important to be able to independently control the modal gain on the edge, so that the bulk does not interfere with the edge's functionality.
In this Letter, we report such a non-Hermitian contrast between the bulk and edge: 
under a uniform pumping throughout the structure, the bulk shows entirely real eigen spectra,
while only the edge shows gain and loss as guaranteed by the topology of the bulk.
We refer to such a contrast as the \textit{topological edge-gain effect}.
Here, the bulk is in the parity-time ($\mathcal{PT}$)-exact phase protected by a topological gap, while the gapless edge states exhibit gain through a thresholdless $\mathcal{PT}$ phase transition \cite{Bender1998,Ozdemir2019,Li170,Luo2018,Makris2008,Guo2009,Ruter2010,Chong2011,Feng2013,Feng2014,Hodaei2014,Cerjan2016,Ge2014}.
We show that such an effect can be realized in any topological systems that consist of two gapped subsystems, regardless of the specific design, lattice, or any additional symmetry.
As an example application, such a scheme points to a new route towards topological lasers \cite{Harari2018,Bandres2018,Sun2019,Secli2019p,Kartashov2019,Zhao2018,Bahari2017,St-Jean2017,Pilozzi2016,RZYao2018,Longhi2018,Parto2018,Ota2018,RevModPhys.91.015006}, since our approach is explicitly reciprocal, in contrast to existing theoretical models underlying recent experimental developments. 
It demonstrates that topological lasing is intrinsically compatible with reciprocity and hence with standard integrated photonic platforms. 
Furthermore, since modal gain only occurs at the edge under uniform pumping, the selective pumping required in previous works is relieved in our scheme. 

It should be emphasized that this work is demonstrating the edge-gain effect for high dimensional systems and the physics differs from the edge-gain effect in a 1D structure \cite{Hu2011,Parto2018,Yuce2018c,Schomerus2013}. 
In the latter, the edge is essentially one point and is not associated with any edge transport or robustness to edge truncation. In contrast, the edge-gain effect in higher dimensions encloses richer physics related to transport and $\mathcal{PT}$ symmetry, such as robustness to edge defects, superluminal and zero-group velocities, exceptional points, etc.
These features are important for many applications such as robust lasers \cite{Khanikaev2017,RevModPhys.91.015006,Lu2016,Lu2014}.
Moreover, the scheme to realize the 1D edge gain does not apply to higher dimensions either \cite{Hu2011,Esaki2011}. 
Our approach here is conceptually novel, and the result is of interest in a broader range of applications.

\begin{figure}[t]
    \centering
    \includegraphics[width=\linewidth]{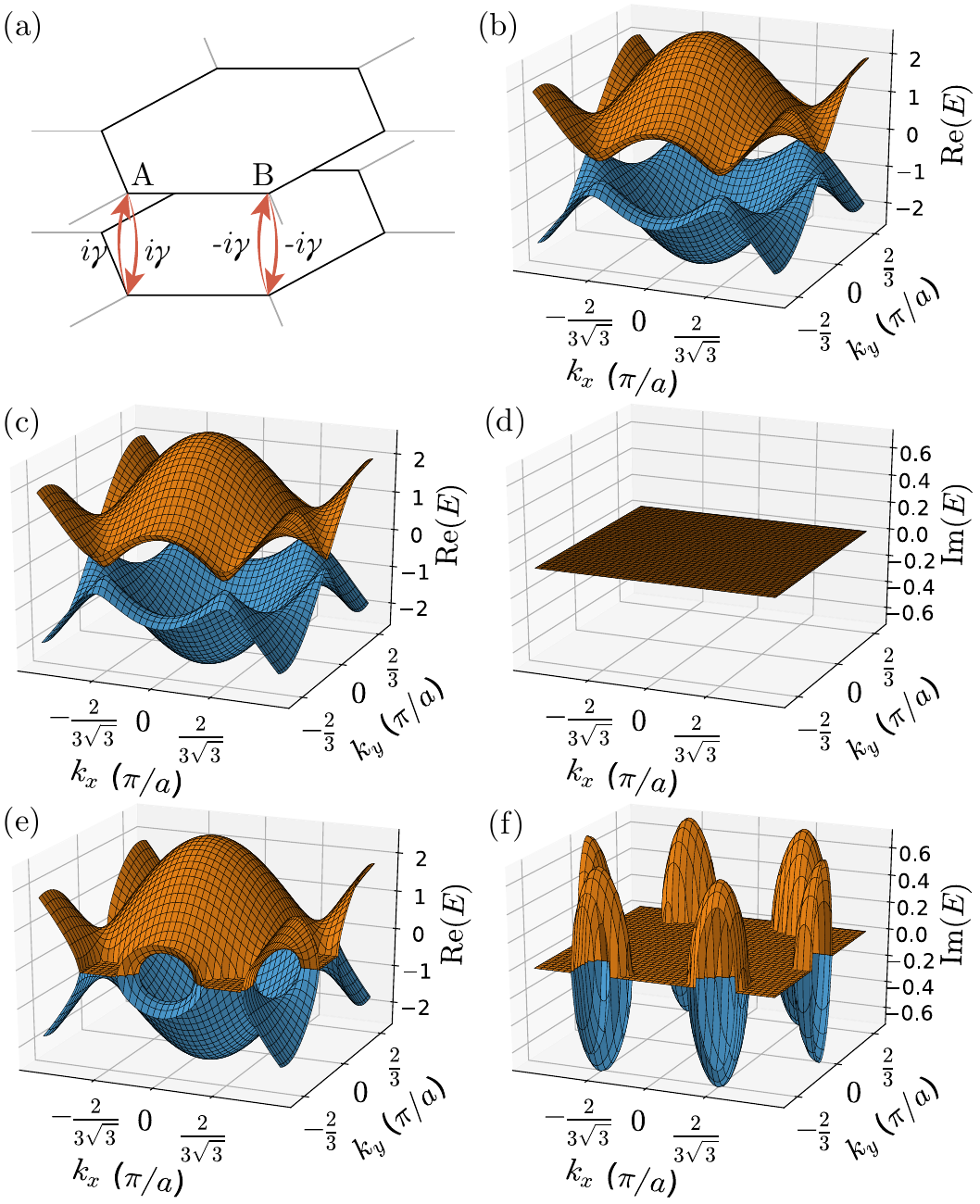}\\
    \caption{
    (a) A schematic of the non-Hermitian QSH model. For convenience, we draw the lattices of the two spins in two layers. The added non-Hermitian couplings are indicated by the red arrows.
    The calculated real (b, c, and e) and imaginary (d and f) parts of the bulk band structure are shown. Here the nearest and next nearest neighbor coupling strengths are set to $t=1$ and $\lambda_{\rm{so}}=0.05$, respectively. $\gamma=0$ for (b),  $\gamma=0.9\times 3\sqrt{3}\lambda_{\rm{so}}$ for (c) and (d), and $\gamma=3.0\times 3\sqrt{3}\lambda_{\rm{so}}$ for (e) and (f).
    }
    \label{fig_bulk_bands}
\end{figure}

\textit{Model.}- To illustrate the basic idea, we consider a concrete model as illustrated by the following Hamiltonian:
\begin{equation}
    H = t\sum_{\braket{ij}} c^{\dagger}_i c_j + i \lambda_{\rm{so}}\sum_{\llangle ij\rrangle} v_{ij} c^\dagger_i s^z c_j +i\gamma \sum_i  \rho_{i} c^\dagger_i s^x c_i.
    \label{eq_H_realspace}
\end{equation}
This Hamiltonian describes interacting spins on a honeycomb lattice, as is depicted in Fig.~\ref{fig_bulk_bands}a. It is based on the Kane-Mele quantum spin-Hall (QSH) model \cite{Kane2005a, Kane2005}. 
The first and the  second terms in Eq.~(\ref{eq_H_realspace}) are the nearest and second nearest neighbor couplings in the Kane-Mele model, where $t$ and $\lambda_{\rm{so}}$ are the respective coupling strengths.
$c_i = (c_{i\uparrow}, c_{i,\downarrow})$ is the annihilation operator for the two spins on site $i$. 
$v_{ij} = \pm1$, depending on the orientation of the two nearest neighbor bonds going from $i$ to $j$. The Pauli matrices $s^i$ operate on the spin subspace.
The third term in Eq.~(\ref{eq_H_realspace}), which is added in our model, describes a non-Hermitian coupling between the spins \cite{Malzard2015}.
Here $\gamma$ is the coupling strength, $\rho_i=\pm 1$, depending on the sub-lattice index. The non-Hermitian coupling is introduced on each site uniformly.

In the momentum space, the Hamiltonian in Eq.~(\ref{eq_H_realspace}) is block diagonalized as
\begin{equation}
    h = d_1\Gamma^1 + d_{12}\Gamma^{12} + d_{15}\Gamma^{15} + i\gamma \Gamma^{13}=
    \begin{pmatrix}
        h_\uparrow &  i\gamma \sigma^z\\
        i\gamma \sigma^z  & h_\downarrow
    \end{pmatrix},
    \label{eq_H_kspace}
\end{equation}
where the first three terms form the original  Kane-Mele Hamiltonian, i.e. $h_{KM} = d_1\Gamma^1 + d_{12}\Gamma^{12} + d_{15}\Gamma^{15}$. 
They generate the diagonal blocks in Eq.~(\ref{eq_H_kspace}), where $h_{\uparrow,\downarrow}$ are the uncoupled Haldane Hamiltonians for each spin. We use the same representation for the Dirac $\Gamma$ matrices as in Ref.~\onlinecite{Kane2005}, which is reproduced here: $\Gamma^{(1,2,3,4,5)}=(\sigma^x\otimes I,\sigma^z\otimes I, \sigma^y\otimes s^x, \sigma^y\otimes s^y, \sigma^y\otimes s^z)$, and the commutators are defined as $\Gamma^{ab}=[\Gamma^a,\Gamma^b]/(2i)$. Here $\sigma^i$ operates in the sublattice space. It follows that $\Gamma^{12}=-\sigma^y\otimes I$, $\Gamma^{15}=\sigma^z\otimes s^z$, and $\Gamma^{13}=\sigma^z\otimes s^x$. 
The expressions for the coefficients are $d_1=t(1+2\cos k_x/2 \cos \sqrt{3}k_y/2)$, $d_{12}=-2t\cos k_x/2\sin \sqrt{3}k_y/2$, $d_{15} = \lambda_{\rm{so}}(2 \sin k_x - 4\sin k_x/2 \cos \sqrt{3}k_y/2)$, where we have set the nearest neighbor distance to 1 \cite{Kane2005}. 
The fourth term in Eq.~(\ref{eq_H_kspace}) is the non-Hermitian addition to the Kane-Mele Hamiltonian. It corresponds to the off-diagonal blocks in the matrix of Eq.~(\ref{eq_H_kspace}).

To understand the effect of the non-Hermitian couplings, we first note that with a zero coupling strength $\gamma$, our model reduces to that of the Kane-Mele. 
There, the two spins are decoupled, and the eigen spectrum is doubly degenerate with a bulk topological gap, as is shown in Fig.~\ref{fig_bulk_bands}b.
The added term residing on the off-diagonal blocks in the matrix of Eq.~(\ref{eq_H_kspace}) couples the two spins.
However, it can be shown that such a non-Hermitian coupling does not couple the two degenerate bulk states on the same side of the gap \footnote{Please see the Supplemental Material for more information, which includes Refs.~\onlinecite{Haus1984}}.
Instead, it only couples two states of different spins on the opposite sides of the gap.
After we switch on the non-Hermitian coupling, the system preserves a $\mathcal{PT}$ symmetry, where $\mathcal{P}=\sigma_x \otimes I$  
is the spatial inversion, and $\mathcal{T} = I \otimes s_x K_0$ is the Bosonic time reversal operator with $K_0$ being the complex conjugation.
To see the implication of this, in Fig.~\ref{fig_bulk_bands}c-f we plot the energy bands after we switch on the non-Hermitian coupling.
With a small coupling strength of $\gamma <3\sqrt{3}\lambda_{\rm{so}}$, the bulk gap remains, and the entire system is in the $\mathcal{PT}$-exact phase, where the eigenenergies of all the bulk bands are real.
With increased coupling strength $\gamma$, the energies of the modes at the upper and lower gap edge approach each other. 
With a larger coupling of $\gamma >3\sqrt{3}\lambda_{\rm{so}}$, the bulk gap closes as the eigenenergies of the upper and lower bands become complex conjugate pairs in the vicinity of $K$ and $K'$ points enclosed by exceptional rings.
The system bulk thus undergoes a $\mathcal{PT}$ phase transition, where the eigenstates exhibit gain and loss.

\begin{figure}[t]
    \centering
    \includegraphics[width=\linewidth]{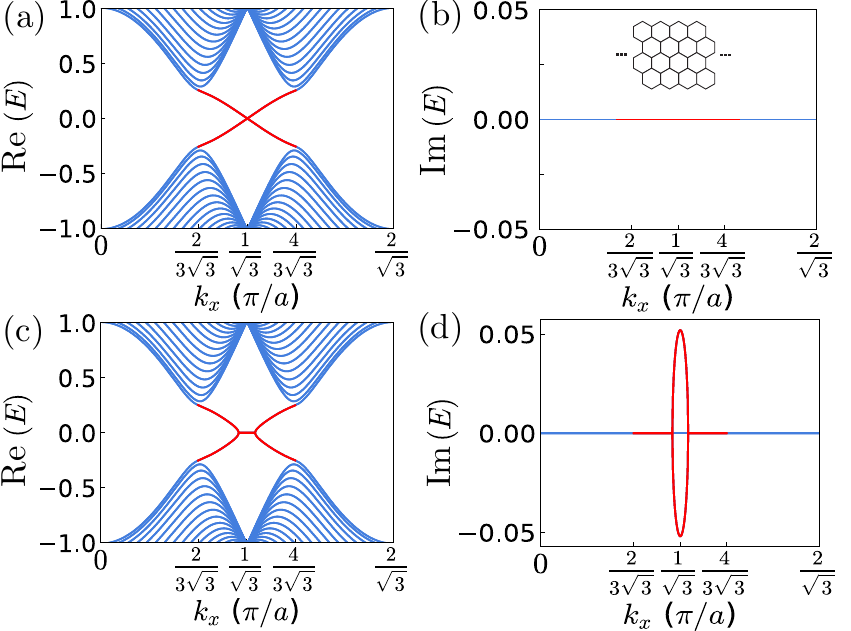}\\
    \caption{
    Calculated real (a and c) and imaginary (b and d) part of the bands in a stripe with zigzag boundary. A schematic of the shape is shown in the inset of (b), with 30 hexagonal cells in the finite direction. $\gamma = 0$ for (a) and (b), and $\gamma = 0.2\times 3\sqrt{3}\lambda_{\rm{so}}$ for (c) and (d). The edge bands are colored in red while those of the bulk are in blue.
    }
    \label{fig_edge_bands}
\end{figure}

The situation at the edge is qualitatively different from that of the bulk. In Fig.~\ref{fig_edge_bands} we show the calculated band structure of this model in a semi-infinite stripe geometry with the zigzag boundary. 
Before the non-Hermitian couplings are applied, the eigen spectrum reduces to that of the original Kane-Mele model featuring a pair of gapless edge states, as is shown in Fig.~\ref{fig_edge_bands}a. 
For a non-Hermitian coupling of $\gamma < 3\sqrt{3} \lambda_{\rm{so}}$, the bulk bands are in the $\mathcal{PT}$-exact phase without gain or loss due to the nonzero bulk gap, as discussed above. 
However, the gapless edge states can always enter the $\mathcal{PT}$-broken phase starting from the degenerate crossing point, in a region between two exceptional points, as is shown in Fig.~\ref{fig_edge_bands}c and d.
Thus, modal gain and loss only appear on the edge but not in the bulk, showing the topological edge-gain effect.
Dynamics of the edge bands in the vicinity of the crossing point can be studied with an effective $2\times 2$ Hamiltonian, and some details are given in the Supplemental Material \cite{Note1}.

\begin{figure}[t]
    \centering
    \includegraphics[width=\linewidth]{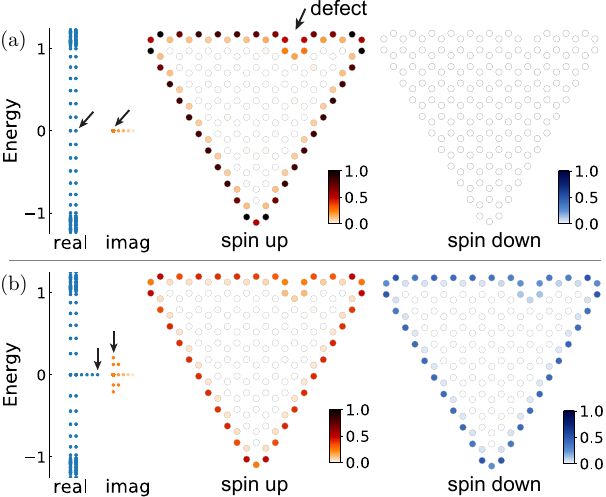}\\
    \caption{
    Calculated wave functions of the model in Eq.~(\ref{eq_H_realspace}) in a finite structure.
    (a) The eigenenergy spectrum without non-Hermitian couplings, and the wave function of the spin-up midgap edge state. The structure is a triangle with a defect on an edge.
    Degenerate energies in the spectrum are offset for visual clarity. 
    The arrows mark the edge state that is plotted.
    The up-spin and down-spin components of the edge state are shown on the right.
    (b) The eigenenergy spectrum with non-Hermitian coupling and the wave function of the edge state with gain. Parameters used for this simulation: $t=1$, $\lambda_{\rm{so}}=0.2$, $\gamma=0.2\times 3\sqrt{3}\lambda_{\rm{so}}$.
    }
    \label{fig_finite}
\end{figure}

In our model, the coupled edge states inherits the robustness against disorder from the edge states in a regular QSH system \cite{Kane2005a,Kane2005,BHZ2006,Konig2007}. In the latter, a defect introduced on the edge will not cause reflection or localization as long as it does not flip the spins. This can be seen in Fig.~\ref{fig_finite}a, where the edge state is extensively distributed in spite of a missing site on the edge.
In our model, the non-Hermitian interaction induces a coupling between the spins. 
However, since the non-Hermitian coupling is uniformly distributed over the entire structure, the eigenstate consists of an admixture of the up and down spin eigenstates of a Hermitian QSH system.  
Consequently, the edge states of our non-Hermitian system carry over the robustness against order, as is observed in Fig.~\ref{fig_finite}b.

Such robustness against disorder has been exploited in the design of topological lasers, as it may help with suppressing the modal competition for better single-mode performance \cite{Harari2018,Bandres2018,Sun2019,Secli2019p,Kartashov2019,Zhao2018,Bahari2017,St-Jean2017,Pilozzi2016,RZYao2018,Longhi2018,Parto2018,Ota2018,RevModPhys.91.015006}.
However, previous approaches to topological lasers are based on non-reciprocal models, which is challenging to implement in optoelectronic platforms.
In contrast, our construction explores topological effects in a time-reversal invariant setting, which is compatible with standard photonic platforms such as photonic crystals or coupled ring resonator arrays.
Moreover, with a uniform pumping everywhere in our system, only the edge states have gain and thus can lase, while the entire bulk has zero gain.
This is in contrast to previous works on topological lasers, where a spatially selective pumping scheme is usually required to ensure that only the edge state lases.

\begin{figure}[t]
    \includegraphics[width=\linewidth]{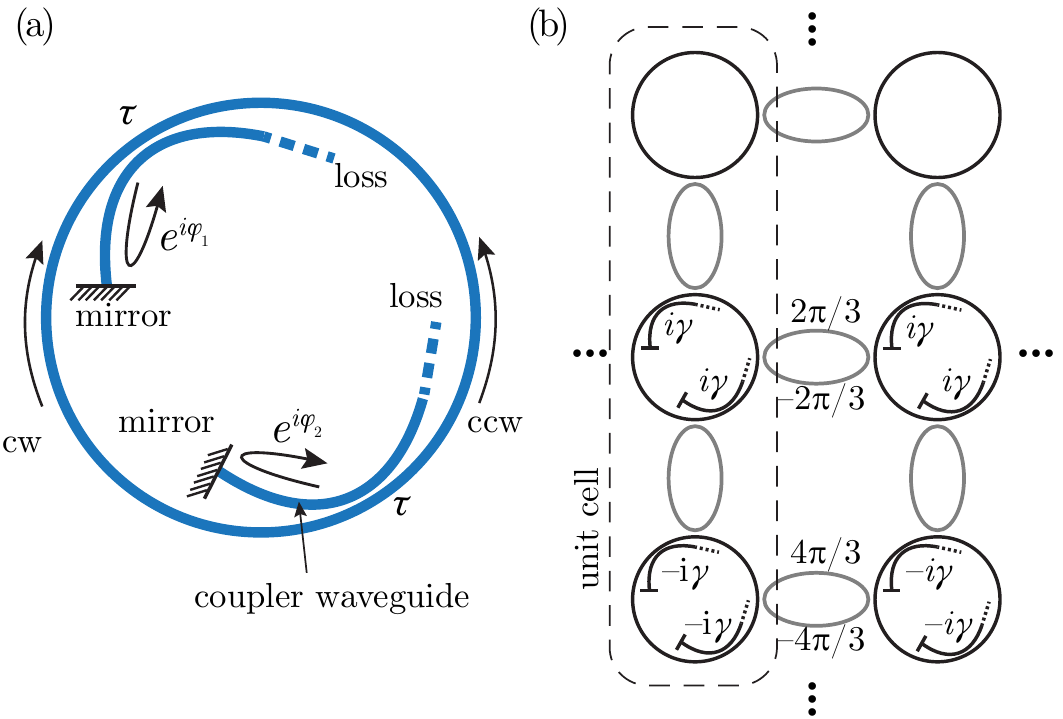}\\
    \caption{Schematic of the coupled ring resonator arrays.
    (a) A ring resonator with two coupler waveguides inside. Each coupler waveguide has a lossy port on one end, and is reflecting on the other end. $1/\tau$ is the leakage rate into the couplers from the ring. $\phi_{1,2}$ are the phases accumulated by waves traveling from the coupling region to the mirror and back, which can be adjusted by changing the waveguide length.
    (b) A lattice of coupled ring resonator arrays creating a QSH system with effective magnetic flux of $1/3$. The lattice structure is identical to that of Ref.~\cite{Hafezi2011} except for the added waveguide couplers inside several the ring resonators. The effective coupling induced by each coupler is marked aside the couplers. The effective phase of the linking rings for ccw waves are also marked on the linking rings.
    }
    \label{fig_lattice}
\end{figure}

\textit{General recipe.}- The topological edge-gain effect can be constructed in any QSH-like systems containing four bands or more, without the need of any additional symmetries such as time reversal or spatial inversion \cite{Note1}.
For a QSH system with time reversal and inversion symmetries,
denoting $h_\uparrow(k)$ and $h_\downarrow(k)$ as the Hamiltonians of the two spins in the momentum space, the following form of a non-Hermitian Hamiltonian
\begin{equation}
    H =
    \begin{pmatrix}
        h_{\uparrow}(k)               & f_1(k) \kappa(k) \\
        f_2(k)\kappa(k) &  h_{\downarrow}(k)
    \end{pmatrix},
    \ \ \
    \kappa(k) = h_\uparrow(k) - h_\downarrow(k),
    \label{eq_recipie}
\end{equation}
provides the topological edge-gain effect, with arbitrary complex functions of $f_{1}(k)$ and $f_{2}(k)$, given that $f_1(k)f_2(k)$ is real.
It can be shown that such a Hamiltonian satisfies 
pseudo-Hermiticity, which is considered as a generalization of $\mathcal{PT}$ symmetry \cite{Note1,Mostafazadeh2002III,Zhang2020,Nixon2016,Siegl2009}.
With a non-Hermitian coupling strength that does not close the bulk gap, the bulk states remain 
without gain or loss, whereas the edge states always exhibit gain thorough a thresholdless phase transition. 
In our example model above, $\kappa = 2d_{15}(k)s^z$, and we have chosen $f_1(k)=f_2(k)=i\gamma/d_{15}(k)$.

Several well-known QSH models do have inversion symmetry, including the Kane-Mele model discussed above, the Bernevig-Hughes-Zhang model discussed in the Supplemental Material \cite{BHZ2006,Konig2007, Note1}, and the model of magnetic flux on a lattice discussed in the following.

\textit{Implementation by coupled ring optical resonators.}-
Here we discuss an implementation of our model in coupled-ring optical resonator arrays. The structure is shown in Fig.~\ref{fig_lattice}, which is based on that studied in Refs.~\onlinecite{Hafezi2011,Harari2018,Bandres2018}.
Here, the two spins are realized by the clockwise (cw) and counter-clockwise (ccw) modes in each ring.
To induce a non-Hermitian coupling between the two spins, we add two waveguide segments evanescently coupled to the ring resonator shown in Fig.~\ref{fig_lattice}a \cite{Fan2003,WonjooSuh2004}. Each of the waveguides has a reflecting end and a lossy end.
The cw (ccw) mode leak into the couplers, and in the designated coupler it gets reflected by the mirror and fed into ccw (cw), respectively. 
Each of such a ring is described by the following coupled-mode equation \cite{Note1}:
\begin{equation}
    i \frac{d}{dt}
    \begin{pmatrix}
        a_{1}\\a_{2}
    \end{pmatrix}
    =
    \begin{pmatrix}
        - i\frac{2}{\tau}  & i \frac{2}{\tau}e^{i\phi_1}\\
        i \frac{2}{\tau}e^{i\phi_2}  &  -i \frac{2}{\tau}
    \end{pmatrix}
    \begin{pmatrix}
        a_{1}\\a_{2}
    \end{pmatrix},
    \label{eq_coupled_mode_in_ring}
\end{equation}
where $a_{1,2}$ is the wave amplitude of the cw and ccw modes, respectively, $1/\tau$ is the leakage rate into the couplers, and $\phi_{1,2}$ are the phases accumulated by the wave traveling from the coupling region to the mirror and back. $\phi_{1,2}$ can be adjusted by tuning the lengths of the coupling waveguides.
The diagonal terms of $-2i/\tau$ in Eq.~(\ref{eq_coupled_mode_in_ring}) are losses induced by the couplers, which can be compensated by a pumping of the gain medium in each ring resonator. The off-diagonal matrix elements in Eq.~(\ref{eq_coupled_mode_in_ring}) give us the desired non-Hermitian couplings between the cw and ccw modes.

\begin{figure}[b]
    \centering
    \includegraphics[width=\linewidth]{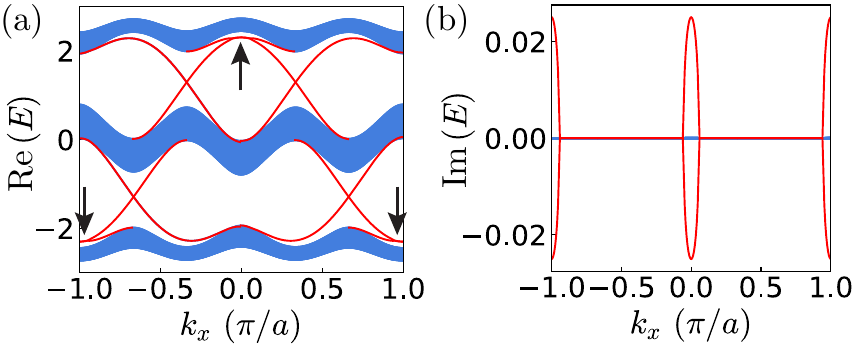}\\
    \caption{Calculated band structure of the QSH system with non-Hermitian couplings in Fig.~\ref{fig_lattice}b. The real and the imaginary parts of the eigenenergies are plotted in (a) and (b), respectively. The bulk bands are colored in blue, while the edge states in red. The location where  edge states enter the $\mathcal{PT}$-broken phase is marked by the black arrows.}
    \label{fig_lattice_bands}
\end{figure}

A lattice of coupled ring resonators with non-Hermitian couplings are shown in Fig.~\ref{fig_lattice}b. 
Apart from the coupler waveguides in each ring, the structure is identical to that used in Ref.~\onlinecite{Hafezi2011}. 
The Hamiltonian of this system takes the form
\begin{equation}
    \begin{aligned}
        H = & -t\sum_{x,y}(c^\dagger_{x+1,y}e^{-i2\pi\alpha y s^z}c_{x,y} + c^\dagger_{x,y+1}c_{x,y} + \rm{H.C.}) \\
        & + \sum_{x,y} c^\dagger_{x,y}  
        \begin{pmatrix}
            0 & i e^{i\varphi}s_{\mu}\gamma^+_{|\mu|} \\
            i e^{-i\varphi}s_{\mu}\gamma^-_{|\mu|} & 0
        \end{pmatrix}
        c_{x,y}.
    \end{aligned}
    \label{eq_mfol}
\end{equation}
Here $t$ is the hopping strength between nearest neighbor modes, $c_{x,y} = (c_{x,y,\uparrow}, c_{x,y,\downarrow})$ is the annihilation operator for the two spins on each site.
The first line of the Hamiltonian describes a square lattice under a uniform magnetic field as described in the Landau gauge.
$\alpha=p/q$ is the magnetic flux through each lattice unit cell, and $p$ and $q$ are incommensurate integers.
The second line in Eq.~(\ref{eq_mfol}) is the non-Hermitian coupling, constructed through the general recipe Eq.~(\ref{eq_recipie}).
Here, $\varphi$ is an arbitrary phase, $\mu = \{[y+(q-1)/2] \bmod q\} - (q-1)/2$, $s_\mu$ is the sign of $\mu$, $\gamma^\pm_0=0$, and $\{\gamma^{+,-}_{1, 2, ... ,(q-1)/2}\}$ is a set of arbitrary real numbers.

We simulate the structure with a magnetic flux of $1/3$, and the photonic lattice is shown in Fig.~\ref{fig_lattice}b.
We adjust the in-ring couplers such that the non-Hermitian coupling coefficients for the three sub-lattice sites are $0$, $i\gamma$, and $-i\gamma$, respectively. 
We calculate the band structure of this system in a stripe geometry that is periodic in $x$ and finite in $y$ with 300 cells. The results are plotted in Fig.~\ref{fig_lattice_bands}. For this calculation, we assume $t=1$, and $\gamma=0.3$, whereas the threshold for the bulk $\mathcal{PT}$ phase transition is found numerically to be $\gamma\approx0.7$.
Again, although pumping is introduced throughout the ring-resonator arrays, gain only manifests on the edge but not in the bulk. 
Thus, same as the Kane-Mele QSH system discussed above, a topological localization of gain to the edge also occurs. 
The coupled ring-resonator array structure in Fig.~\ref{fig_lattice}b can be implemented through the same fabrication process as in previous experimental platforms \cite{Bandres2018, Mittal_Hafezi_2016}.

In recent experimental studies of such a semiconductor ring-resonator setup, the reported topological gap was about 1~nm \cite{Harari2018,Bandres2018,Hodaei2014}. This leads to a 14~cm$^{-1}$ modal gain differential between  edge and the bulk \cite{Note1}.
Furthermore, the edge dominant mode with the highest gain becomes a lossless dark mode in the absence of pumping, whereas all other modes are lossy in this case. 
The fact that this edge mode is decoupled from the lossy ports guarantees the potential for low-threshold lasing. 

In conclusion, we have shown that a uniform non-Hermitian topological material can have complex eigenvalues only in the edge states but not in the bulk.
Such a topological edge-gain effect is protected by the bulk topology and $\mathcal{PT}$ symmetry, which lead to different $\mathcal{PT}$ phases in the bulk and on the edge.
This effect can be generally induced in any topological material that contains two gapped subsystems, with a total of four bands or more.
Such a separation of non-Hermitian phases between the bulk and the edge adds to the understanding of non-Hermitian topology.
Our result provides the gain and loss control on the edge that is independent of  the bulk, thus can be useful in various applications of non-Hermitian topological edge states.
The designing of topological lasers can also benefit from this effect. In contrast to existing theoretical models for topological lasers, our construction is explicitly reciprocal. Furthermore, as gain is topologically localized to the edge, a non-degenerate lasing state is defined globally without the need of selective pumping.

\begin{acknowledgments}
    This work is supported in part by an U. S. Air Force Office of Scientific Research (AFOSR) Multidisciplinary University Research Initiative (MURI) project (Grant No. FA9550-17-1-0002), 
    a Vannevar Bush Faculty Fellowship from U. S. Department of Defense (DOD) (Grant No. N00014-17-1-3030), 
    U. S. Department of Defense Joint Directed Energy Transition Office (DE-JTO) (Grant No. N00014-17-1-2557), 
    and Office of High Energy Physics of U. S. Department of Energy (DOE) Office of Science (contract No. de-sc0019380).
\end{acknowledgments}


\begin{thebibliography}{89}%
    \makeatletter
    \providecommand \@ifxundefined [1]{%
     \@ifx{#1\undefined}
    }%
    \providecommand \@ifnum [1]{%
     \ifnum #1\expandafter \@firstoftwo
     \else \expandafter \@secondoftwo
     \fi
    }%
    \providecommand \@ifx [1]{%
     \ifx #1\expandafter \@firstoftwo
     \else \expandafter \@secondoftwo
     \fi
    }%
    \providecommand \natexlab [1]{#1}%
    \providecommand \enquote  [1]{``#1''}%
    \providecommand \bibnamefont  [1]{#1}%
    \providecommand \bibfnamefont [1]{#1}%
    \providecommand \citenamefont [1]{#1}%
    \providecommand \href@noop [0]{\@secondoftwo}%
    \providecommand \href [0]{\begingroup \@sanitize@url \@href}%
    \providecommand \@href[1]{\@@startlink{#1}\@@href}%
    \providecommand \@@href[1]{\endgroup#1\@@endlink}%
    \providecommand \@sanitize@url [0]{\catcode `\\12\catcode `\$12\catcode
      `\&12\catcode `\#12\catcode `\^12\catcode `\_12\catcode `\%12\relax}%
    \providecommand \@@startlink[1]{}%
    \providecommand \@@endlink[0]{}%
    \providecommand \url  [0]{\begingroup\@sanitize@url \@url }%
    \providecommand \@url [1]{\endgroup\@href {#1}{\urlprefix }}%
    \providecommand \urlprefix  [0]{URL }%
    \providecommand \Eprint [0]{\href }%
    \providecommand \doibase [0]{http://dx.doi.org/}%
    \providecommand \selectlanguage [0]{\@gobble}%
    \providecommand \bibinfo  [0]{\@secondoftwo}%
    \providecommand \bibfield  [0]{\@secondoftwo}%
    \providecommand \translation [1]{[#1]}%
    \providecommand \BibitemOpen [0]{}%
    \providecommand \bibitemStop [0]{}%
    \providecommand \bibitemNoStop [0]{.\EOS\space}%
    \providecommand \EOS [0]{\spacefactor3000\relax}%
    \providecommand \BibitemShut  [1]{\csname bibitem#1\endcsname}%
    \let\auto@bib@innerbib\@empty
    \bibitem [{\citenamefont {Qi}\ and\ \citenamefont {Zhang}(2011)}]{Qi2011}%
      \BibitemOpen
      \bibfield  {author} {\bibinfo {author} {\bibfnamefont {X.-L.}\ \bibnamefont
      {Qi}}\ and\ \bibinfo {author} {\bibfnamefont {S.-C.}\ \bibnamefont {Zhang}},\
      }\href {\doibase 10.1103/RevModPhys.83.1057} {\bibfield  {journal} {\bibinfo
      {journal} {Reviews of Modern Physics}\ }\textbf {\bibinfo {volume} {83}},\
      \bibinfo {pages} {1057} (\bibinfo {year} {2011})}\BibitemShut {NoStop}%
    \bibitem [{\citenamefont {Hasan}\ and\ \citenamefont {Kane}(2010)}]{Hasan2010}%
      \BibitemOpen
      \bibfield  {author} {\bibinfo {author} {\bibfnamefont {M.~Z.}\ \bibnamefont
      {Hasan}}\ and\ \bibinfo {author} {\bibfnamefont {C.~L.}\ \bibnamefont
      {Kane}},\ }\href {\doibase 10.1103/RevModPhys.82.3045} {\bibfield  {journal}
      {\bibinfo  {journal} {Reviews of Modern Physics}\ }\textbf {\bibinfo {volume}
      {82}},\ \bibinfo {pages} {3045} (\bibinfo {year} {2010})}\BibitemShut
      {NoStop}%
    \bibitem [{\citenamefont {Maciejko}\ \emph {et~al.}(2011)\citenamefont
      {Maciejko}, \citenamefont {Hughes},\ and\ \citenamefont
      {Zhang}}]{Maciejko2011}%
      \BibitemOpen
      \bibfield  {author} {\bibinfo {author} {\bibfnamefont {J.}~\bibnamefont
      {Maciejko}}, \bibinfo {author} {\bibfnamefont {T.~L.}\ \bibnamefont
      {Hughes}}, \ and\ \bibinfo {author} {\bibfnamefont {S.-C.}\ \bibnamefont
      {Zhang}},\ }\href {\doibase 10.1146/annurev-conmatphys-062910-140538}
      {\bibfield  {journal} {\bibinfo  {journal} {Annual Review of Condensed Matter
      Physics}\ }\textbf {\bibinfo {volume} {2}},\ \bibinfo {pages} {31} (\bibinfo
      {year} {2011})}\BibitemShut {NoStop}%
    \bibitem [{\citenamefont {Kane}\ and\ \citenamefont
      {Mele}(2005{\natexlab{a}})}]{Kane2005a}%
      \BibitemOpen
      \bibfield  {author} {\bibinfo {author} {\bibfnamefont {C.~L.}\ \bibnamefont
      {Kane}}\ and\ \bibinfo {author} {\bibfnamefont {E.~J.}\ \bibnamefont
      {Mele}},\ }\href {\doibase 10.1103/PhysRevLett.95.226801} {\bibfield
      {journal} {\bibinfo  {journal} {Physical Review Letters}\ }\textbf {\bibinfo
      {volume} {95}},\ \bibinfo {pages} {226801} (\bibinfo {year}
      {2005}{\natexlab{a}})}\BibitemShut {NoStop}%
    \bibitem [{\citenamefont {Kane}\ and\ \citenamefont
      {Mele}(2005{\natexlab{b}})}]{Kane2005}%
      \BibitemOpen
      \bibfield  {author} {\bibinfo {author} {\bibfnamefont {C.~L.}\ \bibnamefont
      {Kane}}\ and\ \bibinfo {author} {\bibfnamefont {E.~J.}\ \bibnamefont
      {Mele}},\ }\href {\doibase 10.1103/PhysRevLett.95.146802} {\bibfield
      {journal} {\bibinfo  {journal} {Physical Review Letters}\ }\textbf {\bibinfo
      {volume} {95}},\ \bibinfo {pages} {146802} (\bibinfo {year}
      {2005}{\natexlab{b}})}\BibitemShut {NoStop}%
    \bibitem [{\citenamefont {Bernevig}\ \emph {et~al.}(2006)\citenamefont
      {Bernevig}, \citenamefont {Hughes},\ and\ \citenamefont {Zhang}}]{BHZ2006}%
      \BibitemOpen
      \bibfield  {author} {\bibinfo {author} {\bibfnamefont {B.~A.}\ \bibnamefont
      {Bernevig}}, \bibinfo {author} {\bibfnamefont {T.~L.}\ \bibnamefont
      {Hughes}}, \ and\ \bibinfo {author} {\bibfnamefont {S.-C.}\ \bibnamefont
      {Zhang}},\ }\href {\doibase 10.1126/science.1133734} {\bibfield  {journal}
      {\bibinfo  {journal} {Science}\ }\textbf {\bibinfo {volume} {314}},\ \bibinfo
      {pages} {1757} (\bibinfo {year} {2006})}\BibitemShut {NoStop}%
    \bibitem [{\citenamefont {Konig}\ \emph {et~al.}(2007)\citenamefont {Konig},
      \citenamefont {Wiedmann}, \citenamefont {Brune}, \citenamefont {Roth},
      \citenamefont {Buhmann}, \citenamefont {Molenkamp}, \citenamefont {Qi},\ and\
      \citenamefont {Zhang}}]{Konig2007}%
      \BibitemOpen
      \bibfield  {author} {\bibinfo {author} {\bibfnamefont {M.}~\bibnamefont
      {Konig}}, \bibinfo {author} {\bibfnamefont {S.}~\bibnamefont {Wiedmann}},
      \bibinfo {author} {\bibfnamefont {C.}~\bibnamefont {Brune}}, \bibinfo
      {author} {\bibfnamefont {A.}~\bibnamefont {Roth}}, \bibinfo {author}
      {\bibfnamefont {H.}~\bibnamefont {Buhmann}}, \bibinfo {author} {\bibfnamefont
      {L.~W.}\ \bibnamefont {Molenkamp}}, \bibinfo {author} {\bibfnamefont {X.-L.}\
      \bibnamefont {Qi}}, \ and\ \bibinfo {author} {\bibfnamefont {S.-C.}\
      \bibnamefont {Zhang}},\ }\href {\doibase 10.1126/science.1148047} {\bibfield
      {journal} {\bibinfo  {journal} {Science}\ }\textbf {\bibinfo {volume}
      {318}},\ \bibinfo {pages} {766} (\bibinfo {year} {2007})}\BibitemShut
      {NoStop}%
    \bibitem [{\citenamefont {Haldane}(1988)}]{Haldane1988}%
      \BibitemOpen
      \bibfield  {author} {\bibinfo {author} {\bibfnamefont {F.~D.~M.}\
      \bibnamefont {Haldane}},\ }\href {\doibase 10.1103/PhysRevLett.61.2015}
      {\bibfield  {journal} {\bibinfo  {journal} {Physical Review Letters}\
      }\textbf {\bibinfo {volume} {61}},\ \bibinfo {pages} {2015} (\bibinfo {year}
      {1988})}\BibitemShut {NoStop}%
    \bibitem [{\citenamefont {Thouless}\ \emph {et~al.}(1982)\citenamefont
      {Thouless}, \citenamefont {Kohmoto}, \citenamefont {Nightingale},\ and\
      \citenamefont {den Nijs}}]{Thouless1982}%
      \BibitemOpen
      \bibfield  {author} {\bibinfo {author} {\bibfnamefont {D.~J.}\ \bibnamefont
      {Thouless}}, \bibinfo {author} {\bibfnamefont {M.}~\bibnamefont {Kohmoto}},
      \bibinfo {author} {\bibfnamefont {M.~P.}\ \bibnamefont {Nightingale}}, \ and\
      \bibinfo {author} {\bibfnamefont {M.}~\bibnamefont {den Nijs}},\ }\href
      {\doibase 10.1103/PhysRevLett.49.405} {\bibfield  {journal} {\bibinfo
      {journal} {Physical Review Letters}\ }\textbf {\bibinfo {volume} {49}},\
      \bibinfo {pages} {405} (\bibinfo {year} {1982})}\BibitemShut {NoStop}%
    \bibitem [{\citenamefont {Hatsugai}(1993)}]{Hatsugai1993}%
      \BibitemOpen
      \bibfield  {author} {\bibinfo {author} {\bibfnamefont {Y.}~\bibnamefont
      {Hatsugai}},\ }\href {\doibase 10.1103/PhysRevLett.71.3697} {\bibfield
      {journal} {\bibinfo  {journal} {Physical Review Letters}\ }\textbf {\bibinfo
      {volume} {71}},\ \bibinfo {pages} {3697} (\bibinfo {year}
      {1993})}\BibitemShut {NoStop}%
    \bibitem [{\citenamefont {Khanikaev}\ and\ \citenamefont
      {Shvets}(2017)}]{Khanikaev2017}%
      \BibitemOpen
      \bibfield  {author} {\bibinfo {author} {\bibfnamefont {A.~B.}\ \bibnamefont
      {Khanikaev}}\ and\ \bibinfo {author} {\bibfnamefont {G.}~\bibnamefont
      {Shvets}},\ }\href {\doibase 10.1038/s41566-017-0048-5} {\bibfield  {journal}
      {\bibinfo  {journal} {Nature Photonics}\ }\textbf {\bibinfo {volume} {11}},\
      \bibinfo {pages} {763} (\bibinfo {year} {2017})}\BibitemShut {NoStop}%
    \bibitem [{\citenamefont {Ozawa}\ \emph {et~al.}(2019)\citenamefont {Ozawa},
      \citenamefont {Price}, \citenamefont {Amo}, \citenamefont {Goldman},
      \citenamefont {Hafezi}, \citenamefont {Lu}, \citenamefont {Rechtsman},
      \citenamefont {Schuster}, \citenamefont {Simon}, \citenamefont {Zilberberg},\
      and\ \citenamefont {Carusotto}}]{RevModPhys.91.015006}%
      \BibitemOpen
      \bibfield  {author} {\bibinfo {author} {\bibfnamefont {T.}~\bibnamefont
      {Ozawa}}, \bibinfo {author} {\bibfnamefont {H.~M.}\ \bibnamefont {Price}},
      \bibinfo {author} {\bibfnamefont {A.}~\bibnamefont {Amo}}, \bibinfo {author}
      {\bibfnamefont {N.}~\bibnamefont {Goldman}}, \bibinfo {author} {\bibfnamefont
      {M.}~\bibnamefont {Hafezi}}, \bibinfo {author} {\bibfnamefont
      {L.}~\bibnamefont {Lu}}, \bibinfo {author} {\bibfnamefont {M.~C.}\
      \bibnamefont {Rechtsman}}, \bibinfo {author} {\bibfnamefont {D.}~\bibnamefont
      {Schuster}}, \bibinfo {author} {\bibfnamefont {J.}~\bibnamefont {Simon}},
      \bibinfo {author} {\bibfnamefont {O.}~\bibnamefont {Zilberberg}}, \ and\
      \bibinfo {author} {\bibfnamefont {I.}~\bibnamefont {Carusotto}},\ }\href
      {\doibase 10.1103/RevModPhys.91.015006} {\bibfield  {journal} {\bibinfo
      {journal} {Rev. Mod. Phys.}\ }\textbf {\bibinfo {volume} {91}},\ \bibinfo
      {pages} {015006} (\bibinfo {year} {2019})}\BibitemShut {NoStop}%
    \bibitem [{\citenamefont {Lu}\ \emph {et~al.}(2016)\citenamefont {Lu},
      \citenamefont {Joannopoulos},\ and\ \citenamefont
      {Solja{\v{c}}i{\'{c}}}}]{Lu2016}%
      \BibitemOpen
      \bibfield  {author} {\bibinfo {author} {\bibfnamefont {L.}~\bibnamefont
      {Lu}}, \bibinfo {author} {\bibfnamefont {J.~D.}\ \bibnamefont
      {Joannopoulos}}, \ and\ \bibinfo {author} {\bibfnamefont {M.}~\bibnamefont
      {Solja{\v{c}}i{\'{c}}}},\ }\href {\doibase 10.1038/nphys3796} {\bibfield
      {journal} {\bibinfo  {journal} {Nature Physics}\ }\textbf {\bibinfo {volume}
      {12}},\ \bibinfo {pages} {626} (\bibinfo {year} {2016})}\BibitemShut
      {NoStop}%
    \bibitem [{\citenamefont {Lu}\ \emph {et~al.}(2014)\citenamefont {Lu},
      \citenamefont {Joannopoulos},\ and\ \citenamefont
      {Solja{\v{c}}i{\'{c}}}}]{Lu2014}%
      \BibitemOpen
      \bibfield  {author} {\bibinfo {author} {\bibfnamefont {L.}~\bibnamefont
      {Lu}}, \bibinfo {author} {\bibfnamefont {J.~D.}\ \bibnamefont
      {Joannopoulos}}, \ and\ \bibinfo {author} {\bibfnamefont {M.}~\bibnamefont
      {Solja{\v{c}}i{\'{c}}}},\ }\href {\doibase 10.1038/nphoton.2014.248}
      {\bibfield  {journal} {\bibinfo  {journal} {Nature Photonics}\ }\textbf
      {\bibinfo {volume} {8}},\ \bibinfo {pages} {821} (\bibinfo {year}
      {2014})}\BibitemShut {NoStop}%
    \bibitem [{\citenamefont {Liang}\ and\ \citenamefont
      {Chong}(2013)}]{Liang2013a}%
      \BibitemOpen
      \bibfield  {author} {\bibinfo {author} {\bibfnamefont {G.~Q.}\ \bibnamefont
      {Liang}}\ and\ \bibinfo {author} {\bibfnamefont {Y.~D.}\ \bibnamefont
      {Chong}},\ }\href {\doibase 10.1103/PhysRevLett.110.203904} {\bibfield
      {journal} {\bibinfo  {journal} {Physical Review Letters}\ }\textbf {\bibinfo
      {volume} {110}},\ \bibinfo {pages} {203904} (\bibinfo {year}
      {2013})}\BibitemShut {NoStop}%
    \bibitem [{\citenamefont {Hafezi}\ \emph {et~al.}(2011)\citenamefont {Hafezi},
      \citenamefont {Demler}, \citenamefont {Lukin},\ and\ \citenamefont
      {Taylor}}]{Hafezi2011}%
      \BibitemOpen
      \bibfield  {author} {\bibinfo {author} {\bibfnamefont {M.}~\bibnamefont
      {Hafezi}}, \bibinfo {author} {\bibfnamefont {E.~A.}\ \bibnamefont {Demler}},
      \bibinfo {author} {\bibfnamefont {M.~D.}\ \bibnamefont {Lukin}}, \ and\
      \bibinfo {author} {\bibfnamefont {J.~M.}\ \bibnamefont {Taylor}},\ }\href
      {\doibase 10.1038/nphys2063} {\bibfield  {journal} {\bibinfo  {journal}
      {Nature Physics}\ }\textbf {\bibinfo {volume} {7}},\ \bibinfo {pages} {907}
      (\bibinfo {year} {2011})}\BibitemShut {NoStop}%
    \bibitem [{\citenamefont {Hafezi}\ \emph {et~al.}(2013)\citenamefont {Hafezi},
      \citenamefont {Mittal}, \citenamefont {Fan}, \citenamefont {Migdall},\ and\
      \citenamefont {Taylor}}]{Hafezi2013}%
      \BibitemOpen
      \bibfield  {author} {\bibinfo {author} {\bibfnamefont {M.}~\bibnamefont
      {Hafezi}}, \bibinfo {author} {\bibfnamefont {S.}~\bibnamefont {Mittal}},
      \bibinfo {author} {\bibfnamefont {J.}~\bibnamefont {Fan}}, \bibinfo {author}
      {\bibfnamefont {A.}~\bibnamefont {Migdall}}, \ and\ \bibinfo {author}
      {\bibfnamefont {J.~M.}\ \bibnamefont {Taylor}},\ }\href {\doibase
      10.1038/nphoton.2013.274} {\bibfield  {journal} {\bibinfo  {journal} {Nature
      Photonics}\ }\textbf {\bibinfo {volume} {7}},\ \bibinfo {pages} {1001}
      (\bibinfo {year} {2013})}\BibitemShut {NoStop}%
    \bibitem [{\citenamefont {Mittal}\ \emph {et~al.}(2016)\citenamefont {Mittal},
      \citenamefont {Ganeshan}, \citenamefont {Fan}, \citenamefont {Vaezi},\ and\
      \citenamefont {Hafezi}}]{Mittal_Hafezi_2016}%
      \BibitemOpen
      \bibfield  {author} {\bibinfo {author} {\bibfnamefont {S.}~\bibnamefont
      {Mittal}}, \bibinfo {author} {\bibfnamefont {S.}~\bibnamefont {Ganeshan}},
      \bibinfo {author} {\bibfnamefont {J.}~\bibnamefont {Fan}}, \bibinfo {author}
      {\bibfnamefont {A.}~\bibnamefont {Vaezi}}, \ and\ \bibinfo {author}
      {\bibfnamefont {M.}~\bibnamefont {Hafezi}},\ }\href {\doibase
      10.1038/nphoton.2016.10} {\bibfield  {journal} {\bibinfo  {journal} {Nature
      Photonics}\ }\textbf {\bibinfo {volume} {10}},\ \bibinfo {pages} {180}
      (\bibinfo {year} {2016})}\BibitemShut {NoStop}%
    \bibitem [{\citenamefont {Yang}\ \emph {et~al.}(2019)\citenamefont {Yang},
      \citenamefont {Gao}, \citenamefont {Xue}, \citenamefont {Zhang},
      \citenamefont {He}, \citenamefont {Yang}, \citenamefont {Singh},
      \citenamefont {Chong}, \citenamefont {Zhang},\ and\ \citenamefont
      {Chen}}]{Yang2019}%
      \BibitemOpen
      \bibfield  {author} {\bibinfo {author} {\bibfnamefont {Y.}~\bibnamefont
      {Yang}}, \bibinfo {author} {\bibfnamefont {Z.}~\bibnamefont {Gao}}, \bibinfo
      {author} {\bibfnamefont {H.}~\bibnamefont {Xue}}, \bibinfo {author}
      {\bibfnamefont {L.}~\bibnamefont {Zhang}}, \bibinfo {author} {\bibfnamefont
      {M.}~\bibnamefont {He}}, \bibinfo {author} {\bibfnamefont {Z.}~\bibnamefont
      {Yang}}, \bibinfo {author} {\bibfnamefont {R.}~\bibnamefont {Singh}},
      \bibinfo {author} {\bibfnamefont {Y.}~\bibnamefont {Chong}}, \bibinfo
      {author} {\bibfnamefont {B.}~\bibnamefont {Zhang}}, \ and\ \bibinfo {author}
      {\bibfnamefont {H.}~\bibnamefont {Chen}},\ }\href {\doibase
      10.1038/s41586-018-0829-0} {\bibfield  {journal} {\bibinfo  {journal}
      {Nature}\ }\textbf {\bibinfo {volume} {565}},\ \bibinfo {pages} {622}
      (\bibinfo {year} {2019})}\BibitemShut {NoStop}%
    \bibitem [{\citenamefont {Lin}\ \emph {et~al.}(2018)\citenamefont {Lin},
      \citenamefont {Sun}, \citenamefont {Xiao}, \citenamefont {Zhang},\ and\
      \citenamefont {Fan}}]{Lin2018}%
      \BibitemOpen
      \bibfield  {author} {\bibinfo {author} {\bibfnamefont {Q.}~\bibnamefont
      {Lin}}, \bibinfo {author} {\bibfnamefont {X.-Q.}\ \bibnamefont {Sun}},
      \bibinfo {author} {\bibfnamefont {M.}~\bibnamefont {Xiao}}, \bibinfo {author}
      {\bibfnamefont {S.-C.}\ \bibnamefont {Zhang}}, \ and\ \bibinfo {author}
      {\bibfnamefont {S.}~\bibnamefont {Fan}},\ }\href {\doibase
      10.1126/sciadv.aat2774} {\bibfield  {journal} {\bibinfo  {journal} {Science
      Advances}\ }\textbf {\bibinfo {volume} {4}},\ \bibinfo {pages} {eaat2774}
      (\bibinfo {year} {2018})}\BibitemShut {NoStop}%
    \bibitem [{\citenamefont {Nalitov}\ \emph {et~al.}(2015)\citenamefont
      {Nalitov}, \citenamefont {Malpuech}, \citenamefont {Ter{\c{c}}as},\ and\
      \citenamefont {Solnyshkov}}]{Nalitov2015}%
      \BibitemOpen
      \bibfield  {author} {\bibinfo {author} {\bibfnamefont {A.~V.}\ \bibnamefont
      {Nalitov}}, \bibinfo {author} {\bibfnamefont {G.}~\bibnamefont {Malpuech}},
      \bibinfo {author} {\bibfnamefont {H.}~\bibnamefont {Ter{\c{c}}as}}, \ and\
      \bibinfo {author} {\bibfnamefont {D.~D.}\ \bibnamefont {Solnyshkov}},\ }\href
      {\doibase 10.1103/PhysRevLett.114.026803} {\bibfield  {journal} {\bibinfo
      {journal} {Physical Review Letters}\ }\textbf {\bibinfo {volume} {114}},\
      \bibinfo {pages} {026803} (\bibinfo {year} {2015})}\BibitemShut {NoStop}%
    \bibitem [{\citenamefont {Kort-Kamp}(2017)}]{Kort-Kamp2017}%
      \BibitemOpen
      \bibfield  {author} {\bibinfo {author} {\bibfnamefont {W.~J.~M.}\
      \bibnamefont {Kort-Kamp}},\ }\href {\doibase 10.1103/PhysRevLett.119.147401}
      {\bibfield  {journal} {\bibinfo  {journal} {Physical Review Letters}\
      }\textbf {\bibinfo {volume} {119}},\ \bibinfo {pages} {147401} (\bibinfo
      {year} {2017})}\BibitemShut {NoStop}%
    \bibitem [{\citenamefont {Rechtsman}\ \emph {et~al.}(2013)\citenamefont
      {Rechtsman}, \citenamefont {Zeuner}, \citenamefont {Plotnik}, \citenamefont
      {Lumer}, \citenamefont {Podolsky}, \citenamefont {Dreisow}, \citenamefont
      {Nolte}, \citenamefont {Segev},\ and\ \citenamefont
      {Szameit}}]{Rechtsman2013}%
      \BibitemOpen
      \bibfield  {author} {\bibinfo {author} {\bibfnamefont {M.~C.}\ \bibnamefont
      {Rechtsman}}, \bibinfo {author} {\bibfnamefont {J.~M.}\ \bibnamefont
      {Zeuner}}, \bibinfo {author} {\bibfnamefont {Y.}~\bibnamefont {Plotnik}},
      \bibinfo {author} {\bibfnamefont {Y.}~\bibnamefont {Lumer}}, \bibinfo
      {author} {\bibfnamefont {D.}~\bibnamefont {Podolsky}}, \bibinfo {author}
      {\bibfnamefont {F.}~\bibnamefont {Dreisow}}, \bibinfo {author} {\bibfnamefont
      {S.}~\bibnamefont {Nolte}}, \bibinfo {author} {\bibfnamefont
      {M.}~\bibnamefont {Segev}}, \ and\ \bibinfo {author} {\bibfnamefont
      {A.}~\bibnamefont {Szameit}},\ }\href {\doibase 10.1038/nature12066}
      {\bibfield  {journal} {\bibinfo  {journal} {Nature}\ }\textbf {\bibinfo
      {volume} {496}},\ \bibinfo {pages} {196} (\bibinfo {year}
      {2013})}\BibitemShut {NoStop}%
    \bibitem [{\citenamefont {Khanikaev}\ \emph {et~al.}(2013)\citenamefont
      {Khanikaev}, \citenamefont {{Hossein Mousavi}}, \citenamefont {Tse},
      \citenamefont {Kargarian}, \citenamefont {MacDonald},\ and\ \citenamefont
      {Shvets}}]{Khanikaev2013}%
      \BibitemOpen
      \bibfield  {author} {\bibinfo {author} {\bibfnamefont {A.~B.}\ \bibnamefont
      {Khanikaev}}, \bibinfo {author} {\bibfnamefont {S.}~\bibnamefont {{Hossein
      Mousavi}}}, \bibinfo {author} {\bibfnamefont {W.-K.}\ \bibnamefont {Tse}},
      \bibinfo {author} {\bibfnamefont {M.}~\bibnamefont {Kargarian}}, \bibinfo
      {author} {\bibfnamefont {A.~H.}\ \bibnamefont {MacDonald}}, \ and\ \bibinfo
      {author} {\bibfnamefont {G.}~\bibnamefont {Shvets}},\ }\href {\doibase
      10.1038/nmat3520} {\bibfield  {journal} {\bibinfo  {journal} {Nature
      Materials}\ }\textbf {\bibinfo {volume} {12}},\ \bibinfo {pages} {233}
      (\bibinfo {year} {2013})}\BibitemShut {NoStop}%
    \bibitem [{\citenamefont {Bliokh}\ \emph {et~al.}(2015)\citenamefont {Bliokh},
      \citenamefont {Smirnova},\ and\ \citenamefont {Nori}}]{Bliokh2015}%
      \BibitemOpen
      \bibfield  {author} {\bibinfo {author} {\bibfnamefont {K.~Y.}\ \bibnamefont
      {Bliokh}}, \bibinfo {author} {\bibfnamefont {D.}~\bibnamefont {Smirnova}}, \
      and\ \bibinfo {author} {\bibfnamefont {F.}~\bibnamefont {Nori}},\ }\href
      {\doibase 10.1126/science.aaa9519} {\bibfield  {journal} {\bibinfo  {journal}
      {Science}\ }\textbf {\bibinfo {volume} {348}},\ \bibinfo {pages} {1448}
      (\bibinfo {year} {2015})}\BibitemShut {NoStop}%
    \bibitem [{\citenamefont {Onoda}\ \emph {et~al.}(2004)\citenamefont {Onoda},
      \citenamefont {Murakami},\ and\ \citenamefont {Nagaosa}}]{Onoda2004}%
      \BibitemOpen
      \bibfield  {author} {\bibinfo {author} {\bibfnamefont {M.}~\bibnamefont
      {Onoda}}, \bibinfo {author} {\bibfnamefont {S.}~\bibnamefont {Murakami}}, \
      and\ \bibinfo {author} {\bibfnamefont {N.}~\bibnamefont {Nagaosa}},\ }\href
      {\doibase 10.1103/PhysRevLett.93.083901} {\bibfield  {journal} {\bibinfo
      {journal} {Physical Review Letters}\ }\textbf {\bibinfo {volume} {93}},\
      \bibinfo {pages} {083901} (\bibinfo {year} {2004})}\BibitemShut {NoStop}%
    \bibitem [{\citenamefont {Wang}\ \emph {et~al.}(2009)\citenamefont {Wang},
      \citenamefont {Chong}, \citenamefont {Joannopoulos},\ and\ \citenamefont
      {Solja{\v{c}}i{\'{c}}}}]{Wang2009}%
      \BibitemOpen
      \bibfield  {author} {\bibinfo {author} {\bibfnamefont {Z.}~\bibnamefont
      {Wang}}, \bibinfo {author} {\bibfnamefont {Y.}~\bibnamefont {Chong}},
      \bibinfo {author} {\bibfnamefont {J.~D.}\ \bibnamefont {Joannopoulos}}, \
      and\ \bibinfo {author} {\bibfnamefont {M.}~\bibnamefont
      {Solja{\v{c}}i{\'{c}}}},\ }\href {\doibase 10.1038/nature08293} {\bibfield
      {journal} {\bibinfo  {journal} {Nature}\ }\textbf {\bibinfo {volume} {461}},\
      \bibinfo {pages} {772} (\bibinfo {year} {2009})}\BibitemShut {NoStop}%
    \bibitem [{\citenamefont {Barik}\ \emph {et~al.}(2018)\citenamefont {Barik},
      \citenamefont {Karasahin}, \citenamefont {Flower}, \citenamefont {Cai},
      \citenamefont {Miyake}, \citenamefont {DeGottardi}, \citenamefont {Hafezi},\
      and\ \citenamefont {Waks}}]{Barik2018}%
      \BibitemOpen
      \bibfield  {author} {\bibinfo {author} {\bibfnamefont {S.}~\bibnamefont
      {Barik}}, \bibinfo {author} {\bibfnamefont {A.}~\bibnamefont {Karasahin}},
      \bibinfo {author} {\bibfnamefont {C.}~\bibnamefont {Flower}}, \bibinfo
      {author} {\bibfnamefont {T.}~\bibnamefont {Cai}}, \bibinfo {author}
      {\bibfnamefont {H.}~\bibnamefont {Miyake}}, \bibinfo {author} {\bibfnamefont
      {W.}~\bibnamefont {DeGottardi}}, \bibinfo {author} {\bibfnamefont
      {M.}~\bibnamefont {Hafezi}}, \ and\ \bibinfo {author} {\bibfnamefont
      {E.}~\bibnamefont {Waks}},\ }\href {\doibase 10.1126/science.aaq0327}
      {\bibfield  {journal} {\bibinfo  {journal} {Science}\ }\textbf {\bibinfo
      {volume} {359}},\ \bibinfo {pages} {666} (\bibinfo {year}
      {2018})}\BibitemShut {NoStop}%
    \bibitem [{\citenamefont {Haldane}\ and\ \citenamefont
      {Raghu}(2008)}]{Haldane2008}%
      \BibitemOpen
      \bibfield  {author} {\bibinfo {author} {\bibfnamefont {F.~D.~M.}\
      \bibnamefont {Haldane}}\ and\ \bibinfo {author} {\bibfnamefont
      {S.}~\bibnamefont {Raghu}},\ }\href {\doibase 10.1103/PhysRevLett.100.013904}
      {\bibfield  {journal} {\bibinfo  {journal} {Physical Review Letters}\
      }\textbf {\bibinfo {volume} {100}},\ \bibinfo {pages} {013904} (\bibinfo
      {year} {2008})}\BibitemShut {NoStop}%
    \bibitem [{\citenamefont {Liu}\ \emph {et~al.}(2019)\citenamefont {Liu},
      \citenamefont {Zhang}, \citenamefont {Ai}, \citenamefont {Gong},
      \citenamefont {Kawabata}, \citenamefont {Ueda},\ and\ \citenamefont
      {Nori}}]{Liu2019}%
      \BibitemOpen
      \bibfield  {author} {\bibinfo {author} {\bibfnamefont {T.}~\bibnamefont
      {Liu}}, \bibinfo {author} {\bibfnamefont {Y.-R.}\ \bibnamefont {Zhang}},
      \bibinfo {author} {\bibfnamefont {Q.}~\bibnamefont {Ai}}, \bibinfo {author}
      {\bibfnamefont {Z.}~\bibnamefont {Gong}}, \bibinfo {author} {\bibfnamefont
      {K.}~\bibnamefont {Kawabata}}, \bibinfo {author} {\bibfnamefont
      {M.}~\bibnamefont {Ueda}}, \ and\ \bibinfo {author} {\bibfnamefont
      {F.}~\bibnamefont {Nori}},\ }\href {\doibase 10.1103/PhysRevLett.122.076801}
      {\bibfield  {journal} {\bibinfo  {journal} {Physical Review Letters}\
      }\textbf {\bibinfo {volume} {122}},\ \bibinfo {pages} {076801} (\bibinfo
      {year} {2019})}\BibitemShut {NoStop}%
    \bibitem [{\citenamefont {Kawabata}\ \emph {et~al.}(2018)\citenamefont
      {Kawabata}, \citenamefont {Shiozaki},\ and\ \citenamefont
      {Ueda}}]{Kawabata2018}%
      \BibitemOpen
      \bibfield  {author} {\bibinfo {author} {\bibfnamefont {K.}~\bibnamefont
      {Kawabata}}, \bibinfo {author} {\bibfnamefont {K.}~\bibnamefont {Shiozaki}},
      \ and\ \bibinfo {author} {\bibfnamefont {M.}~\bibnamefont {Ueda}},\ }\href
      {\doibase 10.1103/PhysRevB.98.165148} {\bibfield  {journal} {\bibinfo
      {journal} {Physical Review B}\ }\textbf {\bibinfo {volume} {98}},\ \bibinfo
      {pages} {165148} (\bibinfo {year} {2018})}\BibitemShut {NoStop}%
    \bibitem [{\citenamefont {Leykam}\ \emph {et~al.}(2017)\citenamefont {Leykam},
      \citenamefont {Bliokh}, \citenamefont {Huang}, \citenamefont {Chong},\ and\
      \citenamefont {Nori}}]{Leykam2017}%
      \BibitemOpen
      \bibfield  {author} {\bibinfo {author} {\bibfnamefont {D.}~\bibnamefont
      {Leykam}}, \bibinfo {author} {\bibfnamefont {K.~Y.}\ \bibnamefont {Bliokh}},
      \bibinfo {author} {\bibfnamefont {C.}~\bibnamefont {Huang}}, \bibinfo
      {author} {\bibfnamefont {Y.~D.}\ \bibnamefont {Chong}}, \ and\ \bibinfo
      {author} {\bibfnamefont {F.}~\bibnamefont {Nori}},\ }\href {\doibase
      10.1103/PhysRevLett.118.040401} {\bibfield  {journal} {\bibinfo  {journal}
      {Physical Review Letters}\ }\textbf {\bibinfo {volume} {118}},\ \bibinfo
      {pages} {040401} (\bibinfo {year} {2017})}\BibitemShut {NoStop}%
    \bibitem [{\citenamefont {Zeuner}\ \emph {et~al.}(2015)\citenamefont {Zeuner},
      \citenamefont {Rechtsman}, \citenamefont {Plotnik}, \citenamefont {Lumer},
      \citenamefont {Nolte}, \citenamefont {Rudner}, \citenamefont {Segev},\ and\
      \citenamefont {Szameit}}]{Zeuner2015}%
      \BibitemOpen
      \bibfield  {author} {\bibinfo {author} {\bibfnamefont {J.~M.}\ \bibnamefont
      {Zeuner}}, \bibinfo {author} {\bibfnamefont {M.~C.}\ \bibnamefont
      {Rechtsman}}, \bibinfo {author} {\bibfnamefont {Y.}~\bibnamefont {Plotnik}},
      \bibinfo {author} {\bibfnamefont {Y.}~\bibnamefont {Lumer}}, \bibinfo
      {author} {\bibfnamefont {S.}~\bibnamefont {Nolte}}, \bibinfo {author}
      {\bibfnamefont {M.~S.}\ \bibnamefont {Rudner}}, \bibinfo {author}
      {\bibfnamefont {M.}~\bibnamefont {Segev}}, \ and\ \bibinfo {author}
      {\bibfnamefont {A.}~\bibnamefont {Szameit}},\ }\href {\doibase
      10.1103/PhysRevLett.115.040402} {\bibfield  {journal} {\bibinfo  {journal}
      {Physical Review Letters}\ }\textbf {\bibinfo {volume} {115}},\ \bibinfo
      {pages} {040402} (\bibinfo {year} {2015})}\BibitemShut {NoStop}%
    \bibitem [{\citenamefont {Weimann}\ \emph {et~al.}(2017)\citenamefont
      {Weimann}, \citenamefont {Kremer}, \citenamefont {Plotnik}, \citenamefont
      {Lumer}, \citenamefont {Nolte}, \citenamefont {Makris}, \citenamefont
      {Segev}, \citenamefont {Rechtsman},\ and\ \citenamefont
      {Szameit}}]{Weimann2017}%
      \BibitemOpen
      \bibfield  {author} {\bibinfo {author} {\bibfnamefont {S.}~\bibnamefont
      {Weimann}}, \bibinfo {author} {\bibfnamefont {M.}~\bibnamefont {Kremer}},
      \bibinfo {author} {\bibfnamefont {Y.}~\bibnamefont {Plotnik}}, \bibinfo
      {author} {\bibfnamefont {Y.}~\bibnamefont {Lumer}}, \bibinfo {author}
      {\bibfnamefont {S.}~\bibnamefont {Nolte}}, \bibinfo {author} {\bibfnamefont
      {K.~G.}\ \bibnamefont {Makris}}, \bibinfo {author} {\bibfnamefont
      {M.}~\bibnamefont {Segev}}, \bibinfo {author} {\bibfnamefont {M.~C.}\
      \bibnamefont {Rechtsman}}, \ and\ \bibinfo {author} {\bibfnamefont
      {A.}~\bibnamefont {Szameit}},\ }\href {\doibase 10.1038/nmat4811} {\bibfield
      {journal} {\bibinfo  {journal} {Nature Materials}\ }\textbf {\bibinfo
      {volume} {16}},\ \bibinfo {pages} {433} (\bibinfo {year} {2017})}\BibitemShut
      {NoStop}%
    \bibitem [{\citenamefont {Liang}\ and\ \citenamefont
      {Huang}(2013)}]{Liang2013}%
      \BibitemOpen
      \bibfield  {author} {\bibinfo {author} {\bibfnamefont {S.-D.}\ \bibnamefont
      {Liang}}\ and\ \bibinfo {author} {\bibfnamefont {G.-Y.}\ \bibnamefont
      {Huang}},\ }\href {\doibase 10.1103/PhysRevA.87.012118} {\bibfield  {journal}
      {\bibinfo  {journal} {Physical Review A}\ }\textbf {\bibinfo {volume} {87}},\
      \bibinfo {pages} {012118} (\bibinfo {year} {2013})}\BibitemShut {NoStop}%
    \bibitem [{\citenamefont {Kunst}\ \emph {et~al.}(2018)\citenamefont {Kunst},
      \citenamefont {Edvardsson}, \citenamefont {Budich},\ and\ \citenamefont
      {Bergholtz}}]{Kunst2018}%
      \BibitemOpen
      \bibfield  {author} {\bibinfo {author} {\bibfnamefont {F.~K.}\ \bibnamefont
      {Kunst}}, \bibinfo {author} {\bibfnamefont {E.}~\bibnamefont {Edvardsson}},
      \bibinfo {author} {\bibfnamefont {J.~C.}\ \bibnamefont {Budich}}, \ and\
      \bibinfo {author} {\bibfnamefont {E.~J.}\ \bibnamefont {Bergholtz}},\ }\href
      {\doibase 10.1103/PhysRevLett.121.026808} {\bibfield  {journal} {\bibinfo
      {journal} {Physical Review Letters}\ }\textbf {\bibinfo {volume} {121}},\
      \bibinfo {pages} {026808} (\bibinfo {year} {2018})}\BibitemShut {NoStop}%
    \bibitem [{\citenamefont {Shen}\ \emph {et~al.}(2018)\citenamefont {Shen},
      \citenamefont {Zhen},\ and\ \citenamefont {Fu}}]{Shen2018}%
      \BibitemOpen
      \bibfield  {author} {\bibinfo {author} {\bibfnamefont {H.}~\bibnamefont
      {Shen}}, \bibinfo {author} {\bibfnamefont {B.}~\bibnamefont {Zhen}}, \ and\
      \bibinfo {author} {\bibfnamefont {L.}~\bibnamefont {Fu}},\ }\href {\doibase
      10.1103/PhysRevLett.120.146402} {\bibfield  {journal} {\bibinfo  {journal}
      {Physical Review Letters}\ }\textbf {\bibinfo {volume} {120}},\ \bibinfo
      {pages} {146402} (\bibinfo {year} {2018})}\BibitemShut {NoStop}%
    \bibitem [{\citenamefont {Esaki}\ \emph {et~al.}(2011)\citenamefont {Esaki},
      \citenamefont {Sato}, \citenamefont {Hasebe},\ and\ \citenamefont
      {Kohmoto}}]{Esaki2011}%
      \BibitemOpen
      \bibfield  {author} {\bibinfo {author} {\bibfnamefont {K.}~\bibnamefont
      {Esaki}}, \bibinfo {author} {\bibfnamefont {M.}~\bibnamefont {Sato}},
      \bibinfo {author} {\bibfnamefont {K.}~\bibnamefont {Hasebe}}, \ and\ \bibinfo
      {author} {\bibfnamefont {M.}~\bibnamefont {Kohmoto}},\ }\href {\doibase
      10.1103/PhysRevB.84.205128} {\bibfield  {journal} {\bibinfo  {journal}
      {Physical Review B}\ }\textbf {\bibinfo {volume} {84}},\ \bibinfo {pages}
      {205128} (\bibinfo {year} {2011})}\BibitemShut {NoStop}%
    \bibitem [{\citenamefont {Malzard}\ \emph {et~al.}(2015)\citenamefont
      {Malzard}, \citenamefont {Poli},\ and\ \citenamefont
      {Schomerus}}]{Malzard2015}%
      \BibitemOpen
      \bibfield  {author} {\bibinfo {author} {\bibfnamefont {S.}~\bibnamefont
      {Malzard}}, \bibinfo {author} {\bibfnamefont {C.}~\bibnamefont {Poli}}, \
      and\ \bibinfo {author} {\bibfnamefont {H.}~\bibnamefont {Schomerus}},\ }\href
      {\doibase 10.1103/PhysRevLett.115.200402} {\bibfield  {journal} {\bibinfo
      {journal} {Physical Review Letters}\ }\textbf {\bibinfo {volume} {115}},\
      \bibinfo {pages} {200402} (\bibinfo {year} {2015})}\BibitemShut {NoStop}%
    \bibitem [{\citenamefont {Dangel}\ \emph {et~al.}(2018)\citenamefont {Dangel},
      \citenamefont {Wagner}, \citenamefont {Cartarius}, \citenamefont {Main},\
      and\ \citenamefont {Wunner}}]{Dangel2018}%
      \BibitemOpen
      \bibfield  {author} {\bibinfo {author} {\bibfnamefont {F.}~\bibnamefont
      {Dangel}}, \bibinfo {author} {\bibfnamefont {M.}~\bibnamefont {Wagner}},
      \bibinfo {author} {\bibfnamefont {H.}~\bibnamefont {Cartarius}}, \bibinfo
      {author} {\bibfnamefont {J.}~\bibnamefont {Main}}, \ and\ \bibinfo {author}
      {\bibfnamefont {G.}~\bibnamefont {Wunner}},\ }\href {\doibase
      10.1103/PhysRevA.98.013628} {\bibfield  {journal} {\bibinfo  {journal}
      {Physical Review A}\ }\textbf {\bibinfo {volume} {98}},\ \bibinfo {pages}
      {013628} (\bibinfo {year} {2018})}\BibitemShut {NoStop}%
    \bibitem [{\citenamefont {{Noh}}\ \emph {et~al.}(2018)\citenamefont {{Noh}},
      \citenamefont {{Benalcazar}}, \citenamefont {{Hughes}},\ and\ \citenamefont
      {{Rechtsman}}}]{CLEO_Hughes_Rechtsman_8426563}%
      \BibitemOpen
      \bibfield  {author} {\bibinfo {author} {\bibfnamefont {J.}~\bibnamefont
      {{Noh}}}, \bibinfo {author} {\bibfnamefont {W.~A.}\ \bibnamefont
      {{Benalcazar}}}, \bibinfo {author} {\bibfnamefont {T.~L.}\ \bibnamefont
      {{Hughes}}}, \ and\ \bibinfo {author} {\bibfnamefont {M.~C.}\ \bibnamefont
      {{Rechtsman}}},\ }in\ \href {\doibase 10.1364/CLEO_QELS.2018.FM2Q.2} {\emph
      {\bibinfo {booktitle} {2018 Conference on Lasers and Electro-Optics
      (CLEO)}}},\ \bibinfo {series and number} {\bibinfo {number} {FM2Q.2}}\
      (\bibinfo {year} {2018})\BibitemShut {NoStop}%
    \bibitem [{\citenamefont {Kawabata}\ \emph {et~al.}(2019)\citenamefont
      {Kawabata}, \citenamefont {Higashikawa}, \citenamefont {Gong}, \citenamefont
      {Ashida},\ and\ \citenamefont {Ueda}}]{Kawabata2019a}%
      \BibitemOpen
      \bibfield  {author} {\bibinfo {author} {\bibfnamefont {K.}~\bibnamefont
      {Kawabata}}, \bibinfo {author} {\bibfnamefont {S.}~\bibnamefont
      {Higashikawa}}, \bibinfo {author} {\bibfnamefont {Z.}~\bibnamefont {Gong}},
      \bibinfo {author} {\bibfnamefont {Y.}~\bibnamefont {Ashida}}, \ and\ \bibinfo
      {author} {\bibfnamefont {M.}~\bibnamefont {Ueda}},\ }\href {\doibase
      10.1038/s41467-018-08254-y} {\bibfield  {journal} {\bibinfo  {journal}
      {Nature Communications}\ }\textbf {\bibinfo {volume} {10}},\ \bibinfo {pages}
      {297} (\bibinfo {year} {2019})}\BibitemShut {NoStop}%
    \bibitem [{\citenamefont {Hirsbrunner}\ \emph {et~al.}(2019)\citenamefont
      {Hirsbrunner}, \citenamefont {Philip},\ and\ \citenamefont
      {Gilbert}}]{Hirsbrunner2019p}%
      \BibitemOpen
      \bibfield  {author} {\bibinfo {author} {\bibfnamefont {M.~R.}\ \bibnamefont
      {Hirsbrunner}}, \bibinfo {author} {\bibfnamefont {T.~M.}\ \bibnamefont
      {Philip}}, \ and\ \bibinfo {author} {\bibfnamefont {M.~J.}\ \bibnamefont
      {Gilbert}},\ }\href {\doibase 10.1103/PhysRevB.100.081104} {\bibfield
      {journal} {\bibinfo  {journal} {Physical Review B}\ }\textbf {\bibinfo
      {volume} {100}},\ \bibinfo {pages} {081104} (\bibinfo {year}
      {2019})}\BibitemShut {NoStop}%
    \bibitem [{\citenamefont {Kremer}\ \emph {et~al.}(2019)\citenamefont {Kremer},
      \citenamefont {Biesenthal}, \citenamefont {Maczewsky}, \citenamefont
      {Heinrich}, \citenamefont {Thomale},\ and\ \citenamefont
      {Szameit}}]{Kremer2019}%
      \BibitemOpen
      \bibfield  {author} {\bibinfo {author} {\bibfnamefont {M.}~\bibnamefont
      {Kremer}}, \bibinfo {author} {\bibfnamefont {T.}~\bibnamefont {Biesenthal}},
      \bibinfo {author} {\bibfnamefont {L.~J.}\ \bibnamefont {Maczewsky}}, \bibinfo
      {author} {\bibfnamefont {M.}~\bibnamefont {Heinrich}}, \bibinfo {author}
      {\bibfnamefont {R.}~\bibnamefont {Thomale}}, \ and\ \bibinfo {author}
      {\bibfnamefont {A.}~\bibnamefont {Szameit}},\ }\href {\doibase
      10.1038/s41467-018-08104-x} {\bibfield  {journal} {\bibinfo  {journal}
      {Nature Communications}\ }\textbf {\bibinfo {volume} {10}},\ \bibinfo {pages}
      {435} (\bibinfo {year} {2019})}\BibitemShut {NoStop}%
    \bibitem [{\citenamefont {Gong}\ \emph {et~al.}(2018)\citenamefont {Gong},
      \citenamefont {Ashida}, \citenamefont {Kawabata}, \citenamefont {Takasan},
      \citenamefont {Higashikawa},\ and\ \citenamefont {Ueda}}]{Gong2018}%
      \BibitemOpen
      \bibfield  {author} {\bibinfo {author} {\bibfnamefont {Z.}~\bibnamefont
      {Gong}}, \bibinfo {author} {\bibfnamefont {Y.}~\bibnamefont {Ashida}},
      \bibinfo {author} {\bibfnamefont {K.}~\bibnamefont {Kawabata}}, \bibinfo
      {author} {\bibfnamefont {K.}~\bibnamefont {Takasan}}, \bibinfo {author}
      {\bibfnamefont {S.}~\bibnamefont {Higashikawa}}, \ and\ \bibinfo {author}
      {\bibfnamefont {M.}~\bibnamefont {Ueda}},\ }\href {\doibase
      10.1103/PhysRevX.8.031079} {\bibfield  {journal} {\bibinfo  {journal}
      {Physical Review X}\ }\textbf {\bibinfo {volume} {8}},\ \bibinfo {pages}
      {031079} (\bibinfo {year} {2018})}\BibitemShut {NoStop}%
    \bibitem [{\citenamefont {Yao}\ and\ \citenamefont {Wang}(2018)}]{Yao2018}%
      \BibitemOpen
      \bibfield  {author} {\bibinfo {author} {\bibfnamefont {S.}~\bibnamefont
      {Yao}}\ and\ \bibinfo {author} {\bibfnamefont {Z.}~\bibnamefont {Wang}},\
      }\href {\doibase 10.1103/PhysRevLett.121.086803} {\bibfield  {journal}
      {\bibinfo  {journal} {Physical Review Letters}\ }\textbf {\bibinfo {volume}
      {121}},\ \bibinfo {pages} {086803} (\bibinfo {year} {2018})}\BibitemShut
      {NoStop}%
    \bibitem [{\citenamefont {{Hassani Gangaraj}}\ and\ \citenamefont
      {Monticone}(2018)}]{HassaniGangaraj2018}%
      \BibitemOpen
      \bibfield  {author} {\bibinfo {author} {\bibfnamefont {S.~A.}\ \bibnamefont
      {{Hassani Gangaraj}}}\ and\ \bibinfo {author} {\bibfnamefont
      {F.}~\bibnamefont {Monticone}},\ }\href {\doibase
      10.1103/PhysRevLett.121.093901} {\bibfield  {journal} {\bibinfo  {journal}
      {Physical Review Letters}\ }\textbf {\bibinfo {volume} {121}},\ \bibinfo
      {pages} {093901} (\bibinfo {year} {2018})}\BibitemShut {NoStop}%
    \bibitem [{\citenamefont {Yuce}(2018{\natexlab{a}})}]{Yuce2018}%
      \BibitemOpen
      \bibfield  {author} {\bibinfo {author} {\bibfnamefont {C.}~\bibnamefont
      {Yuce}},\ }\href {\doibase 10.1103/PhysRevA.98.012111} {\bibfield  {journal}
      {\bibinfo  {journal} {Physical Review A}\ }\textbf {\bibinfo {volume} {98}},\
      \bibinfo {pages} {012111} (\bibinfo {year} {2018}{\natexlab{a}})}\BibitemShut
      {NoStop}%
    \bibitem [{\citenamefont {Harari}\ \emph {et~al.}(2018)\citenamefont {Harari},
      \citenamefont {Bandres}, \citenamefont {Lumer}, \citenamefont {Rechtsman},
      \citenamefont {Chong}, \citenamefont {Khajavikhan}, \citenamefont
      {Christodoulides},\ and\ \citenamefont {Segev}}]{Harari2018}%
      \BibitemOpen
      \bibfield  {author} {\bibinfo {author} {\bibfnamefont {G.}~\bibnamefont
      {Harari}}, \bibinfo {author} {\bibfnamefont {M.~A.}\ \bibnamefont {Bandres}},
      \bibinfo {author} {\bibfnamefont {Y.}~\bibnamefont {Lumer}}, \bibinfo
      {author} {\bibfnamefont {M.~C.}\ \bibnamefont {Rechtsman}}, \bibinfo {author}
      {\bibfnamefont {Y.~D.}\ \bibnamefont {Chong}}, \bibinfo {author}
      {\bibfnamefont {M.}~\bibnamefont {Khajavikhan}}, \bibinfo {author}
      {\bibfnamefont {D.~N.}\ \bibnamefont {Christodoulides}}, \ and\ \bibinfo
      {author} {\bibfnamefont {M.}~\bibnamefont {Segev}},\ }\href {\doibase
      10.1126/science.aar4003} {\bibfield  {journal} {\bibinfo  {journal}
      {Science}\ }\textbf {\bibinfo {volume} {359}},\ \bibinfo {pages} {eaar4003}
      (\bibinfo {year} {2018})}\BibitemShut {NoStop}%
    \bibitem [{\citenamefont {Secl{\`{i}}}\ \emph {et~al.}(2019)\citenamefont
      {Secl{\`{i}}}, \citenamefont {Capone},\ and\ \citenamefont
      {Carusotto}}]{Secli2019p}%
      \BibitemOpen
      \bibfield  {author} {\bibinfo {author} {\bibfnamefont {M.}~\bibnamefont
      {Secl{\`{i}}}}, \bibinfo {author} {\bibfnamefont {M.}~\bibnamefont {Capone}},
      \ and\ \bibinfo {author} {\bibfnamefont {I.}~\bibnamefont {Carusotto}},\
      }\href {\doibase 10.1103/PhysRevResearch.1.033148} {\bibfield  {journal}
      {\bibinfo  {journal} {Physical Review Research}\ }\textbf {\bibinfo {volume}
      {1}},\ \bibinfo {pages} {033148} (\bibinfo {year} {2019})}\BibitemShut
      {NoStop}%
    \bibitem [{\citenamefont {Takata}\ and\ \citenamefont
      {Notomi}(2018)}]{Takata2018}%
      \BibitemOpen
      \bibfield  {author} {\bibinfo {author} {\bibfnamefont {K.}~\bibnamefont
      {Takata}}\ and\ \bibinfo {author} {\bibfnamefont {M.}~\bibnamefont
      {Notomi}},\ }\href {\doibase 10.1103/PhysRevLett.121.213902} {\bibfield
      {journal} {\bibinfo  {journal} {Physical Review Letters}\ }\textbf {\bibinfo
      {volume} {121}},\ \bibinfo {pages} {213902} (\bibinfo {year}
      {2018})}\BibitemShut {NoStop}%
    \bibitem [{\citenamefont {Bahari}\ \emph {et~al.}(2017)\citenamefont {Bahari},
      \citenamefont {Ndao}, \citenamefont {Vallini}, \citenamefont {{El Amili}},
      \citenamefont {Fainman},\ and\ \citenamefont {Kant{\'{e}}}}]{Bahari2017}%
      \BibitemOpen
      \bibfield  {author} {\bibinfo {author} {\bibfnamefont {B.}~\bibnamefont
      {Bahari}}, \bibinfo {author} {\bibfnamefont {A.}~\bibnamefont {Ndao}},
      \bibinfo {author} {\bibfnamefont {F.}~\bibnamefont {Vallini}}, \bibinfo
      {author} {\bibfnamefont {A.}~\bibnamefont {{El Amili}}}, \bibinfo {author}
      {\bibfnamefont {Y.}~\bibnamefont {Fainman}}, \ and\ \bibinfo {author}
      {\bibfnamefont {B.}~\bibnamefont {Kant{\'{e}}}},\ }\href {\doibase
      10.1126/science.aao4551} {\bibfield  {journal} {\bibinfo  {journal}
      {Science}\ }\textbf {\bibinfo {volume} {358}},\ \bibinfo {pages} {636}
      (\bibinfo {year} {2017})}\BibitemShut {NoStop}%
    \bibitem [{\citenamefont {Ni}\ \emph {et~al.}(2018)\citenamefont {Ni},
      \citenamefont {Smirnova}, \citenamefont {Poddubny}, \citenamefont {Leykam},
      \citenamefont {Chong},\ and\ \citenamefont {Khanikaev}}]{Ni2018}%
      \BibitemOpen
      \bibfield  {author} {\bibinfo {author} {\bibfnamefont {X.}~\bibnamefont
      {Ni}}, \bibinfo {author} {\bibfnamefont {D.}~\bibnamefont {Smirnova}},
      \bibinfo {author} {\bibfnamefont {A.}~\bibnamefont {Poddubny}}, \bibinfo
      {author} {\bibfnamefont {D.}~\bibnamefont {Leykam}}, \bibinfo {author}
      {\bibfnamefont {Y.}~\bibnamefont {Chong}}, \ and\ \bibinfo {author}
      {\bibfnamefont {A.~B.}\ \bibnamefont {Khanikaev}},\ }\href {\doibase
      10.1103/PhysRevB.98.165129} {\bibfield  {journal} {\bibinfo  {journal}
      {Physical Review B}\ }\textbf {\bibinfo {volume} {98}},\ \bibinfo {pages}
      {165129} (\bibinfo {year} {2018})}\BibitemShut {NoStop}%
    \bibitem [{\citenamefont {Hou}\ \emph {et~al.}(2019)\citenamefont {Hou},
      \citenamefont {Wu},\ and\ \citenamefont {Zhang}}]{Hou2019}%
      \BibitemOpen
      \bibfield  {author} {\bibinfo {author} {\bibfnamefont {J.}~\bibnamefont
      {Hou}}, \bibinfo {author} {\bibfnamefont {Y.-J.}\ \bibnamefont {Wu}}, \ and\
      \bibinfo {author} {\bibfnamefont {C.}~\bibnamefont {Zhang}},\ }\href
      {http://arxiv.org/abs/1910.14606} {\bibfield  {journal} {\bibinfo  {journal}
      {arXiv:1910.14606}\ } (\bibinfo {year} {2019})}\BibitemShut {NoStop}%
    \bibitem [{\citenamefont {Yuce}\ and\ \citenamefont {Oztas}(2018)}]{Yuce2018a}%
      \BibitemOpen
      \bibfield  {author} {\bibinfo {author} {\bibfnamefont {C.}~\bibnamefont
      {Yuce}}\ and\ \bibinfo {author} {\bibfnamefont {Z.}~\bibnamefont {Oztas}},\
      }\href {\doibase 10.1038/s41598-018-35795-5} {\bibfield  {journal} {\bibinfo
      {journal} {Scientific Reports}\ }\textbf {\bibinfo {volume} {8}},\ \bibinfo
      {pages} {17416} (\bibinfo {year} {2018})}\BibitemShut {NoStop}%
    \bibitem [{\citenamefont {Hu}\ and\ \citenamefont {Hughes}(2011)}]{Hu2011}%
      \BibitemOpen
      \bibfield  {author} {\bibinfo {author} {\bibfnamefont {Y.~C.}\ \bibnamefont
      {Hu}}\ and\ \bibinfo {author} {\bibfnamefont {T.~L.}\ \bibnamefont
      {Hughes}},\ }\href {\doibase 10.1103/PhysRevB.84.153101} {\bibfield
      {journal} {\bibinfo  {journal} {Physical Review B}\ }\textbf {\bibinfo
      {volume} {84}},\ \bibinfo {pages} {153101} (\bibinfo {year}
      {2011})}\BibitemShut {NoStop}%
    \bibitem [{\citenamefont {Bender}\ and\ \citenamefont
      {Boettcher}(1998)}]{Bender1998}%
      \BibitemOpen
      \bibfield  {author} {\bibinfo {author} {\bibfnamefont {C.~M.}\ \bibnamefont
      {Bender}}\ and\ \bibinfo {author} {\bibfnamefont {S.}~\bibnamefont
      {Boettcher}},\ }\href {\doibase 10.1103/PhysRevLett.80.5243} {\bibfield
      {journal} {\bibinfo  {journal} {Phys. Rev. Lett.}\ }\textbf {\bibinfo
      {volume} {80}},\ \bibinfo {pages} {5243} (\bibinfo {year}
      {1998})}\BibitemShut {NoStop}%
    \bibitem [{\citenamefont {{\"{O}}zdemir}\ \emph {et~al.}(2019)\citenamefont
      {{\"{O}}zdemir}, \citenamefont {Rotter}, \citenamefont {Nori},\ and\
      \citenamefont {Yang}}]{Ozdemir2019}%
      \BibitemOpen
      \bibfield  {author} {\bibinfo {author} {\bibfnamefont {{\c{S}}.~K.}\
      \bibnamefont {{\"{O}}zdemir}}, \bibinfo {author} {\bibfnamefont
      {S.}~\bibnamefont {Rotter}}, \bibinfo {author} {\bibfnamefont
      {F.}~\bibnamefont {Nori}}, \ and\ \bibinfo {author} {\bibfnamefont
      {L.}~\bibnamefont {Yang}},\ }\href {\doibase 10.1038/s41563-019-0304-9}
      {\bibfield  {journal} {\bibinfo  {journal} {Nature Materials}\ }\textbf
      {\bibinfo {volume} {18}},\ \bibinfo {pages} {783} (\bibinfo {year}
      {2019})}\BibitemShut {NoStop}%
    \bibitem [{\citenamefont {Li}\ \emph {et~al.}(2019)\citenamefont {Li},
      \citenamefont {Peng}, \citenamefont {Han}, \citenamefont {Miri},
      \citenamefont {Li}, \citenamefont {Xiao}, \citenamefont {Zhu}, \citenamefont
      {Zhao}, \citenamefont {Al{\`u}}, \citenamefont {Fan},\ and\ \citenamefont
      {Qiu}}]{Li170}%
      \BibitemOpen
      \bibfield  {author} {\bibinfo {author} {\bibfnamefont {Y.}~\bibnamefont
      {Li}}, \bibinfo {author} {\bibfnamefont {Y.-G.}\ \bibnamefont {Peng}},
      \bibinfo {author} {\bibfnamefont {L.}~\bibnamefont {Han}}, \bibinfo {author}
      {\bibfnamefont {M.-A.}\ \bibnamefont {Miri}}, \bibinfo {author}
      {\bibfnamefont {W.}~\bibnamefont {Li}}, \bibinfo {author} {\bibfnamefont
      {M.}~\bibnamefont {Xiao}}, \bibinfo {author} {\bibfnamefont {X.-F.}\
      \bibnamefont {Zhu}}, \bibinfo {author} {\bibfnamefont {J.}~\bibnamefont
      {Zhao}}, \bibinfo {author} {\bibfnamefont {A.}~\bibnamefont {Al{\`u}}},
      \bibinfo {author} {\bibfnamefont {S.}~\bibnamefont {Fan}}, \ and\ \bibinfo
      {author} {\bibfnamefont {C.-W.}\ \bibnamefont {Qiu}},\ }\href {\doibase
      10.1126/science.aaw6259} {\bibfield  {journal} {\bibinfo  {journal}
      {Science}\ }\textbf {\bibinfo {volume} {364}},\ \bibinfo {pages} {170}
      (\bibinfo {year} {2019})}\BibitemShut {NoStop}%
    \bibitem [{\citenamefont {Luo}\ \emph {et~al.}(2018)\citenamefont {Luo},
      \citenamefont {Li},\ and\ \citenamefont {Lai}}]{Luo2018}%
      \BibitemOpen
      \bibfield  {author} {\bibinfo {author} {\bibfnamefont {J.}~\bibnamefont
      {Luo}}, \bibinfo {author} {\bibfnamefont {J.}~\bibnamefont {Li}}, \ and\
      \bibinfo {author} {\bibfnamefont {Y.}~\bibnamefont {Lai}},\ }\href {\doibase
      10.1103/PhysRevX.8.031035} {\bibfield  {journal} {\bibinfo  {journal}
      {Physical Review X}\ }\textbf {\bibinfo {volume} {8}},\ \bibinfo {pages}
      {031035} (\bibinfo {year} {2018})}\BibitemShut {NoStop}%
    \bibitem [{\citenamefont {Makris}\ \emph {et~al.}(2008)\citenamefont {Makris},
      \citenamefont {El-Ganainy}, \citenamefont {Christodoulides},\ and\
      \citenamefont {Musslimani}}]{Makris2008}%
      \BibitemOpen
      \bibfield  {author} {\bibinfo {author} {\bibfnamefont {K.~G.}\ \bibnamefont
      {Makris}}, \bibinfo {author} {\bibfnamefont {R.}~\bibnamefont {El-Ganainy}},
      \bibinfo {author} {\bibfnamefont {D.~N.}\ \bibnamefont {Christodoulides}}, \
      and\ \bibinfo {author} {\bibfnamefont {Z.~H.}\ \bibnamefont {Musslimani}},\
      }\href {\doibase 10.1103/PhysRevLett.100.103904} {\bibfield  {journal}
      {\bibinfo  {journal} {Physical Review Letters}\ }\textbf {\bibinfo {volume}
      {100}},\ \bibinfo {pages} {103904} (\bibinfo {year} {2008})}\BibitemShut
      {NoStop}%
    \bibitem [{\citenamefont {Guo}\ \emph {et~al.}(2009)\citenamefont {Guo},
      \citenamefont {Salamo}, \citenamefont {Duchesne}, \citenamefont {Morandotti},
      \citenamefont {Volatier-Ravat}, \citenamefont {Aimez}, \citenamefont
      {Siviloglou},\ and\ \citenamefont {Christodoulides}}]{Guo2009}%
      \BibitemOpen
      \bibfield  {author} {\bibinfo {author} {\bibfnamefont {A.}~\bibnamefont
      {Guo}}, \bibinfo {author} {\bibfnamefont {G.~J.}\ \bibnamefont {Salamo}},
      \bibinfo {author} {\bibfnamefont {D.}~\bibnamefont {Duchesne}}, \bibinfo
      {author} {\bibfnamefont {R.}~\bibnamefont {Morandotti}}, \bibinfo {author}
      {\bibfnamefont {M.}~\bibnamefont {Volatier-Ravat}}, \bibinfo {author}
      {\bibfnamefont {V.}~\bibnamefont {Aimez}}, \bibinfo {author} {\bibfnamefont
      {G.~A.}\ \bibnamefont {Siviloglou}}, \ and\ \bibinfo {author} {\bibfnamefont
      {D.~N.}\ \bibnamefont {Christodoulides}},\ }\href {\doibase
      10.1103/PhysRevLett.103.093902} {\bibfield  {journal} {\bibinfo  {journal}
      {Physical Review Letters}\ }\textbf {\bibinfo {volume} {103}},\ \bibinfo
      {pages} {093902} (\bibinfo {year} {2009})}\BibitemShut {NoStop}%
    \bibitem [{\citenamefont {R{\"{u}}ter}\ \emph {et~al.}(2010)\citenamefont
      {R{\"{u}}ter}, \citenamefont {Makris}, \citenamefont {El-Ganainy},
      \citenamefont {Christodoulides}, \citenamefont {Segev},\ and\ \citenamefont
      {Kip}}]{Ruter2010}%
      \BibitemOpen
      \bibfield  {author} {\bibinfo {author} {\bibfnamefont {C.~E.}\ \bibnamefont
      {R{\"{u}}ter}}, \bibinfo {author} {\bibfnamefont {K.~G.}\ \bibnamefont
      {Makris}}, \bibinfo {author} {\bibfnamefont {R.}~\bibnamefont {El-Ganainy}},
      \bibinfo {author} {\bibfnamefont {D.~N.}\ \bibnamefont {Christodoulides}},
      \bibinfo {author} {\bibfnamefont {M.}~\bibnamefont {Segev}}, \ and\ \bibinfo
      {author} {\bibfnamefont {D.}~\bibnamefont {Kip}},\ }\href {\doibase
      10.1038/nphys1515} {\bibfield  {journal} {\bibinfo  {journal} {Nature
      Physics}\ }\textbf {\bibinfo {volume} {6}},\ \bibinfo {pages} {192} (\bibinfo
      {year} {2010})}\BibitemShut {NoStop}%
    \bibitem [{\citenamefont {Chong}\ \emph {et~al.}(2011)\citenamefont {Chong},
      \citenamefont {Ge},\ and\ \citenamefont {Stone}}]{Chong2011}%
      \BibitemOpen
      \bibfield  {author} {\bibinfo {author} {\bibfnamefont {Y.~D.}\ \bibnamefont
      {Chong}}, \bibinfo {author} {\bibfnamefont {L.}~\bibnamefont {Ge}}, \ and\
      \bibinfo {author} {\bibfnamefont {A.~D.}\ \bibnamefont {Stone}},\ }\href
      {\doibase 10.1103/PhysRevLett.106.093902} {\bibfield  {journal} {\bibinfo
      {journal} {Physical Review Letters}\ }\textbf {\bibinfo {volume} {106}},\
      \bibinfo {pages} {093902} (\bibinfo {year} {2011})}\BibitemShut {NoStop}%
    \bibitem [{\citenamefont {Feng}\ \emph {et~al.}(2013)\citenamefont {Feng},
      \citenamefont {Xu}, \citenamefont {Fegadolli}, \citenamefont {Lu},
      \citenamefont {Oliveira}, \citenamefont {Almeida}, \citenamefont {Chen},\
      and\ \citenamefont {Scherer}}]{Feng2013}%
      \BibitemOpen
      \bibfield  {author} {\bibinfo {author} {\bibfnamefont {L.}~\bibnamefont
      {Feng}}, \bibinfo {author} {\bibfnamefont {Y.-L.}\ \bibnamefont {Xu}},
      \bibinfo {author} {\bibfnamefont {W.~S.}\ \bibnamefont {Fegadolli}}, \bibinfo
      {author} {\bibfnamefont {M.-H.}\ \bibnamefont {Lu}}, \bibinfo {author}
      {\bibfnamefont {J.~E.~B.}\ \bibnamefont {Oliveira}}, \bibinfo {author}
      {\bibfnamefont {V.~R.}\ \bibnamefont {Almeida}}, \bibinfo {author}
      {\bibfnamefont {Y.-F.}\ \bibnamefont {Chen}}, \ and\ \bibinfo {author}
      {\bibfnamefont {A.}~\bibnamefont {Scherer}},\ }\href {\doibase
      10.1038/nmat3495} {\bibfield  {journal} {\bibinfo  {journal} {Nature
      Materials}\ }\textbf {\bibinfo {volume} {12}},\ \bibinfo {pages} {108}
      (\bibinfo {year} {2013})}\BibitemShut {NoStop}%
    \bibitem [{\citenamefont {Feng}\ \emph {et~al.}(2014)\citenamefont {Feng},
      \citenamefont {Wong}, \citenamefont {Ma}, \citenamefont {Wang},\ and\
      \citenamefont {Zhang}}]{Feng2014}%
      \BibitemOpen
      \bibfield  {author} {\bibinfo {author} {\bibfnamefont {L.}~\bibnamefont
      {Feng}}, \bibinfo {author} {\bibfnamefont {Z.~J.}\ \bibnamefont {Wong}},
      \bibinfo {author} {\bibfnamefont {R.-M.}\ \bibnamefont {Ma}}, \bibinfo
      {author} {\bibfnamefont {Y.}~\bibnamefont {Wang}}, \ and\ \bibinfo {author}
      {\bibfnamefont {X.}~\bibnamefont {Zhang}},\ }\href {\doibase
      10.1126/science.1258479} {\bibfield  {journal} {\bibinfo  {journal}
      {Science}\ }\textbf {\bibinfo {volume} {346}},\ \bibinfo {pages} {972}
      (\bibinfo {year} {2014})}\BibitemShut {NoStop}%
    \bibitem [{\citenamefont {Hodaei}\ \emph {et~al.}(2014)\citenamefont {Hodaei},
      \citenamefont {Miri}, \citenamefont {Heinrich}, \citenamefont
      {Christodoulides},\ and\ \citenamefont {Khajavikhan}}]{Hodaei2014}%
      \BibitemOpen
      \bibfield  {author} {\bibinfo {author} {\bibfnamefont {H.}~\bibnamefont
      {Hodaei}}, \bibinfo {author} {\bibfnamefont {M.-A.}\ \bibnamefont {Miri}},
      \bibinfo {author} {\bibfnamefont {M.}~\bibnamefont {Heinrich}}, \bibinfo
      {author} {\bibfnamefont {D.~N.}\ \bibnamefont {Christodoulides}}, \ and\
      \bibinfo {author} {\bibfnamefont {M.}~\bibnamefont {Khajavikhan}},\ }\href
      {\doibase 10.1126/science.1258480} {\bibfield  {journal} {\bibinfo  {journal}
      {Science}\ }\textbf {\bibinfo {volume} {346}},\ \bibinfo {pages} {975}
      (\bibinfo {year} {2014})}\BibitemShut {NoStop}%
    \bibitem [{\citenamefont {Cerjan}\ \emph {et~al.}(2016)\citenamefont {Cerjan},
      \citenamefont {Raman},\ and\ \citenamefont {Fan}}]{Cerjan2016}%
      \BibitemOpen
      \bibfield  {author} {\bibinfo {author} {\bibfnamefont {A.}~\bibnamefont
      {Cerjan}}, \bibinfo {author} {\bibfnamefont {A.}~\bibnamefont {Raman}}, \
      and\ \bibinfo {author} {\bibfnamefont {S.}~\bibnamefont {Fan}},\ }\href
      {\doibase 10.1103/PhysRevLett.116.203902} {\bibfield  {journal} {\bibinfo
      {journal} {Physical Review Letters}\ }\textbf {\bibinfo {volume} {116}},\
      \bibinfo {pages} {203902} (\bibinfo {year} {2016})}\BibitemShut {NoStop}%
    \bibitem [{\citenamefont {Ge}\ and\ \citenamefont {Stone}(2014)}]{Ge2014}%
      \BibitemOpen
      \bibfield  {author} {\bibinfo {author} {\bibfnamefont {L.}~\bibnamefont
      {Ge}}\ and\ \bibinfo {author} {\bibfnamefont {A.~D.}\ \bibnamefont {Stone}},\
      }\href {\doibase 10.1103/PhysRevX.4.031011} {\bibfield  {journal} {\bibinfo
      {journal} {Physical Review X}\ }\textbf {\bibinfo {volume} {4}},\ \bibinfo
      {pages} {031011} (\bibinfo {year} {2014})}\BibitemShut {NoStop}%
    \bibitem [{\citenamefont {Bandres}\ \emph {et~al.}(2018)\citenamefont
      {Bandres}, \citenamefont {Wittek}, \citenamefont {Harari}, \citenamefont
      {Parto}, \citenamefont {Ren}, \citenamefont {Segev}, \citenamefont
      {Christodoulides},\ and\ \citenamefont {Khajavikhan}}]{Bandres2018}%
      \BibitemOpen
      \bibfield  {author} {\bibinfo {author} {\bibfnamefont {M.~A.}\ \bibnamefont
      {Bandres}}, \bibinfo {author} {\bibfnamefont {S.}~\bibnamefont {Wittek}},
      \bibinfo {author} {\bibfnamefont {G.}~\bibnamefont {Harari}}, \bibinfo
      {author} {\bibfnamefont {M.}~\bibnamefont {Parto}}, \bibinfo {author}
      {\bibfnamefont {J.}~\bibnamefont {Ren}}, \bibinfo {author} {\bibfnamefont
      {M.}~\bibnamefont {Segev}}, \bibinfo {author} {\bibfnamefont {D.~N.}\
      \bibnamefont {Christodoulides}}, \ and\ \bibinfo {author} {\bibfnamefont
      {M.}~\bibnamefont {Khajavikhan}},\ }\href {\doibase 10.1126/science.aar4005}
      {\bibfield  {journal} {\bibinfo  {journal} {Science}\ }\textbf {\bibinfo
      {volume} {359}},\ \bibinfo {pages} {eaar4005} (\bibinfo {year}
      {2018})}\BibitemShut {NoStop}%
    \bibitem [{\citenamefont {Sun}\ and\ \citenamefont {Hu}(2019)}]{Sun2019}%
      \BibitemOpen
      \bibfield  {author} {\bibinfo {author} {\bibfnamefont {X.-C.}\ \bibnamefont
      {Sun}}\ and\ \bibinfo {author} {\bibfnamefont {X.}~\bibnamefont {Hu}},\
      }\href {http://arxiv.org/abs/1906.02464} {\bibfield  {journal} {\bibinfo
      {journal} {arXiv}\ } (\bibinfo {year} {2019})}\BibitemShut {NoStop}%
    \bibitem [{\citenamefont {Kartashov}\ and\ \citenamefont
      {Skryabin}(2019)}]{Kartashov2019}%
      \BibitemOpen
      \bibfield  {author} {\bibinfo {author} {\bibfnamefont {Y.~V.}\ \bibnamefont
      {Kartashov}}\ and\ \bibinfo {author} {\bibfnamefont {D.~V.}\ \bibnamefont
      {Skryabin}},\ }\href {\doibase 10.1103/PhysRevLett.122.083902} {\bibfield
      {journal} {\bibinfo  {journal} {Physical Review Letters}\ }\textbf {\bibinfo
      {volume} {122}},\ \bibinfo {pages} {083902} (\bibinfo {year}
      {2019})}\BibitemShut {NoStop}%
    \bibitem [{\citenamefont {Zhao}\ \emph {et~al.}(2018)\citenamefont {Zhao},
      \citenamefont {Miao}, \citenamefont {Teimourpour}, \citenamefont {Malzard},
      \citenamefont {El-Ganainy}, \citenamefont {Schomerus},\ and\ \citenamefont
      {Feng}}]{Zhao2018}%
      \BibitemOpen
      \bibfield  {author} {\bibinfo {author} {\bibfnamefont {H.}~\bibnamefont
      {Zhao}}, \bibinfo {author} {\bibfnamefont {P.}~\bibnamefont {Miao}}, \bibinfo
      {author} {\bibfnamefont {M.~H.}\ \bibnamefont {Teimourpour}}, \bibinfo
      {author} {\bibfnamefont {S.}~\bibnamefont {Malzard}}, \bibinfo {author}
      {\bibfnamefont {R.}~\bibnamefont {El-Ganainy}}, \bibinfo {author}
      {\bibfnamefont {H.}~\bibnamefont {Schomerus}}, \ and\ \bibinfo {author}
      {\bibfnamefont {L.}~\bibnamefont {Feng}},\ }\href {\doibase
      10.1038/s41467-018-03434-2} {\bibfield  {journal} {\bibinfo  {journal}
      {Nature Communications}\ }\textbf {\bibinfo {volume} {9}},\ \bibinfo {pages}
      {981} (\bibinfo {year} {2018})}\BibitemShut {NoStop}%
    \bibitem [{\citenamefont {St-Jean}\ \emph {et~al.}(2017)\citenamefont
      {St-Jean}, \citenamefont {Goblot}, \citenamefont {Galopin}, \citenamefont
      {Lema{\^{i}}tre}, \citenamefont {Ozawa}, \citenamefont {{Le Gratiet}},
      \citenamefont {Sagnes}, \citenamefont {Bloch},\ and\ \citenamefont
      {Amo}}]{St-Jean2017}%
      \BibitemOpen
      \bibfield  {author} {\bibinfo {author} {\bibfnamefont {P.}~\bibnamefont
      {St-Jean}}, \bibinfo {author} {\bibfnamefont {V.}~\bibnamefont {Goblot}},
      \bibinfo {author} {\bibfnamefont {E.}~\bibnamefont {Galopin}}, \bibinfo
      {author} {\bibfnamefont {A.}~\bibnamefont {Lema{\^{i}}tre}}, \bibinfo
      {author} {\bibfnamefont {T.}~\bibnamefont {Ozawa}}, \bibinfo {author}
      {\bibfnamefont {L.}~\bibnamefont {{Le Gratiet}}}, \bibinfo {author}
      {\bibfnamefont {I.}~\bibnamefont {Sagnes}}, \bibinfo {author} {\bibfnamefont
      {J.}~\bibnamefont {Bloch}}, \ and\ \bibinfo {author} {\bibfnamefont
      {A.}~\bibnamefont {Amo}},\ }\href {\doibase 10.1038/s41566-017-0006-2}
      {\bibfield  {journal} {\bibinfo  {journal} {Nature Photonics}\ }\textbf
      {\bibinfo {volume} {11}},\ \bibinfo {pages} {651} (\bibinfo {year}
      {2017})}\BibitemShut {NoStop}%
    \bibitem [{\citenamefont {Pilozzi}\ and\ \citenamefont
      {Conti}(2016)}]{Pilozzi2016}%
      \BibitemOpen
      \bibfield  {author} {\bibinfo {author} {\bibfnamefont {L.}~\bibnamefont
      {Pilozzi}}\ and\ \bibinfo {author} {\bibfnamefont {C.}~\bibnamefont
      {Conti}},\ }\href {\doibase 10.1103/PhysRevB.93.195317} {\bibfield  {journal}
      {\bibinfo  {journal} {Physical Review B}\ }\textbf {\bibinfo {volume} {93}},\
      \bibinfo {pages} {195317} (\bibinfo {year} {2016})}\BibitemShut {NoStop}%
    \bibitem [{\citenamefont {Yao}\ \emph {et~al.}(2018)\citenamefont {Yao},
      \citenamefont {Li}, \citenamefont {Zheng}, \citenamefont {An}, \citenamefont
      {Ding}, \citenamefont {Lee}, \citenamefont {Zhang},\ and\ \citenamefont
      {Guo}}]{RZYao2018}%
      \BibitemOpen
      \bibfield  {author} {\bibinfo {author} {\bibfnamefont {R.}~\bibnamefont
      {Yao}}, \bibinfo {author} {\bibfnamefont {H.}~\bibnamefont {Li}}, \bibinfo
      {author} {\bibfnamefont {B.}~\bibnamefont {Zheng}}, \bibinfo {author}
      {\bibfnamefont {S.}~\bibnamefont {An}}, \bibinfo {author} {\bibfnamefont
      {J.}~\bibnamefont {Ding}}, \bibinfo {author} {\bibfnamefont {C.-S.}\
      \bibnamefont {Lee}}, \bibinfo {author} {\bibfnamefont {H.}~\bibnamefont
      {Zhang}}, \ and\ \bibinfo {author} {\bibfnamefont {W.}~\bibnamefont {Guo}},\
      }\href {http://arxiv.org/abs/1804.01587} {\bibfield  {journal} {\bibinfo
      {journal} {arXiv}\ } (\bibinfo {year} {2018})}\BibitemShut {NoStop}%
    \bibitem [{\citenamefont {Longhi}(2018)}]{Longhi2018}%
      \BibitemOpen
      \bibfield  {author} {\bibinfo {author} {\bibfnamefont {S.}~\bibnamefont
      {Longhi}},\ }\href {\doibase 10.1002/andp.201800023} {\bibfield  {journal}
      {\bibinfo  {journal} {Annalen der Physik}\ }\textbf {\bibinfo {volume}
      {530}},\ \bibinfo {pages} {1800023} (\bibinfo {year} {2018})}\BibitemShut
      {NoStop}%
    \bibitem [{\citenamefont {Parto}\ \emph {et~al.}(2018)\citenamefont {Parto},
      \citenamefont {Wittek}, \citenamefont {Hodaei}, \citenamefont {Harari},
      \citenamefont {Bandres}, \citenamefont {Ren}, \citenamefont {Rechtsman},
      \citenamefont {Segev}, \citenamefont {Christodoulides},\ and\ \citenamefont
      {Khajavikhan}}]{Parto2018}%
      \BibitemOpen
      \bibfield  {author} {\bibinfo {author} {\bibfnamefont {M.}~\bibnamefont
      {Parto}}, \bibinfo {author} {\bibfnamefont {S.}~\bibnamefont {Wittek}},
      \bibinfo {author} {\bibfnamefont {H.}~\bibnamefont {Hodaei}}, \bibinfo
      {author} {\bibfnamefont {G.}~\bibnamefont {Harari}}, \bibinfo {author}
      {\bibfnamefont {M.~A.}\ \bibnamefont {Bandres}}, \bibinfo {author}
      {\bibfnamefont {J.}~\bibnamefont {Ren}}, \bibinfo {author} {\bibfnamefont
      {M.~C.}\ \bibnamefont {Rechtsman}}, \bibinfo {author} {\bibfnamefont
      {M.}~\bibnamefont {Segev}}, \bibinfo {author} {\bibfnamefont {D.~N.}\
      \bibnamefont {Christodoulides}}, \ and\ \bibinfo {author} {\bibfnamefont
      {M.}~\bibnamefont {Khajavikhan}},\ }\href {\doibase
      10.1103/PhysRevLett.120.113901} {\bibfield  {journal} {\bibinfo  {journal}
      {Physical Review Letters}\ }\textbf {\bibinfo {volume} {120}},\ \bibinfo
      {pages} {113901} (\bibinfo {year} {2018})}\BibitemShut {NoStop}%
    \bibitem [{\citenamefont {Ota}\ \emph {et~al.}(2018)\citenamefont {Ota},
      \citenamefont {Katsumi}, \citenamefont {Watanabe}, \citenamefont {Iwamoto},\
      and\ \citenamefont {Arakawa}}]{Ota2018}%
      \BibitemOpen
      \bibfield  {author} {\bibinfo {author} {\bibfnamefont {Y.}~\bibnamefont
      {Ota}}, \bibinfo {author} {\bibfnamefont {R.}~\bibnamefont {Katsumi}},
      \bibinfo {author} {\bibfnamefont {K.}~\bibnamefont {Watanabe}}, \bibinfo
      {author} {\bibfnamefont {S.}~\bibnamefont {Iwamoto}}, \ and\ \bibinfo
      {author} {\bibfnamefont {Y.}~\bibnamefont {Arakawa}},\ }\href {\doibase
      10.1038/s42005-018-0083-7} {\bibfield  {journal} {\bibinfo  {journal}
      {Communications Physics}\ }\textbf {\bibinfo {volume} {1}},\ \bibinfo {pages}
      {86} (\bibinfo {year} {2018})}\BibitemShut {NoStop}%
    \bibitem [{\citenamefont {Yuce}(2018{\natexlab{b}})}]{Yuce2018c}%
      \BibitemOpen
      \bibfield  {author} {\bibinfo {author} {\bibfnamefont {C.}~\bibnamefont
      {Yuce}},\ }\href {\doibase 10.1103/PhysRevA.97.042118} {\bibfield  {journal}
      {\bibinfo  {journal} {Physical Review A}\ }\textbf {\bibinfo {volume} {97}},\
      \bibinfo {pages} {042118} (\bibinfo {year} {2018}{\natexlab{b}})}\BibitemShut
      {NoStop}%
    \bibitem [{\citenamefont {Schomerus}(2013)}]{Schomerus2013}%
      \BibitemOpen
      \bibfield  {author} {\bibinfo {author} {\bibfnamefont {H.}~\bibnamefont
      {Schomerus}},\ }\href {\doibase 10.1364/OL.38.001912} {\bibfield  {journal}
      {\bibinfo  {journal} {Optics Letters}\ }\textbf {\bibinfo {volume} {38}},\
      \bibinfo {pages} {1912} (\bibinfo {year} {2013})}\BibitemShut {NoStop}%
    \bibitem [{Note1()}]{Note1}%
      \BibitemOpen
      \bibinfo {note} {Please see the Supplemental Material for more information,
      which includes Refs.~\protect \rev@citealp {Haus1984}}\BibitemShut {NoStop}%
    \bibitem [{\citenamefont {Mostafazadeh}(2002)}]{Mostafazadeh2002III}%
      \BibitemOpen
      \bibfield  {author} {\bibinfo {author} {\bibfnamefont {A.}~\bibnamefont
      {Mostafazadeh}},\ }\href {\doibase 10.1063/1.1489072} {\bibfield  {journal}
      {\bibinfo  {journal} {Journal of Mathematical Physics}\ }\textbf {\bibinfo
      {volume} {43}},\ \bibinfo {pages} {3944} (\bibinfo {year}
      {2002})}\BibitemShut {NoStop}%
    \bibitem [{\citenamefont {Zhang}\ \emph {et~al.}(2020)\citenamefont {Zhang},
      \citenamefont {Qin},\ and\ \citenamefont {Xiao}}]{Zhang2020}%
      \BibitemOpen
      \bibfield  {author} {\bibinfo {author} {\bibfnamefont {R.}~\bibnamefont
      {Zhang}}, \bibinfo {author} {\bibfnamefont {H.}~\bibnamefont {Qin}}, \ and\
      \bibinfo {author} {\bibfnamefont {J.}~\bibnamefont {Xiao}},\ }\href {\doibase
      10.1063/1.5117211} {\bibfield  {journal} {\bibinfo  {journal} {Journal of
      Mathematical Physics}\ }\textbf {\bibinfo {volume} {61}},\ \bibinfo {pages}
      {012101} (\bibinfo {year} {2020})}\BibitemShut {NoStop}%
    \bibitem [{\citenamefont {Nixon}\ and\ \citenamefont {Yang}(2016)}]{Nixon2016}%
      \BibitemOpen
      \bibfield  {author} {\bibinfo {author} {\bibfnamefont {S.}~\bibnamefont
      {Nixon}}\ and\ \bibinfo {author} {\bibfnamefont {J.}~\bibnamefont {Yang}},\
      }\href {\doibase 10.1103/PhysRevA.93.031802} {\bibfield  {journal} {\bibinfo
      {journal} {Physical Review A}\ }\textbf {\bibinfo {volume} {93}},\ \bibinfo
      {pages} {031802} (\bibinfo {year} {2016})}\BibitemShut {NoStop}%
    \bibitem [{\citenamefont {Siegl}(2009)}]{Siegl2009}%
      \BibitemOpen
      \bibfield  {author} {\bibinfo {author} {\bibfnamefont {P.}~\bibnamefont
      {Siegl}},\ }\href {\doibase 10.1007/s12043-009-0119-3} {\bibfield  {journal}
      {\bibinfo  {journal} {Pramana}\ }\textbf {\bibinfo {volume} {73}},\ \bibinfo
      {pages} {279} (\bibinfo {year} {2009})}\BibitemShut {NoStop}%
    \bibitem [{\citenamefont {Fan}\ \emph {et~al.}(2003)\citenamefont {Fan},
      \citenamefont {Suh},\ and\ \citenamefont {Joannopoulos}}]{Fan2003}%
      \BibitemOpen
      \bibfield  {author} {\bibinfo {author} {\bibfnamefont {S.}~\bibnamefont
      {Fan}}, \bibinfo {author} {\bibfnamefont {W.}~\bibnamefont {Suh}}, \ and\
      \bibinfo {author} {\bibfnamefont {J.~D.}\ \bibnamefont {Joannopoulos}},\
      }\href {\doibase 10.1364/JOSAA.20.000569} {\bibfield  {journal} {\bibinfo
      {journal} {Journal of the Optical Society of America A}\ }\textbf {\bibinfo
      {volume} {20}},\ \bibinfo {pages} {569} (\bibinfo {year} {2003})}\BibitemShut
      {NoStop}%
    \bibitem [{\citenamefont {{Wonjoo Suh}}\ \emph {et~al.}(2004)\citenamefont
      {{Wonjoo Suh}}, \citenamefont {{Zheng Wang}},\ and\ \citenamefont {{Shanhui
      Fan}}}]{WonjooSuh2004}%
      \BibitemOpen
      \bibfield  {author} {\bibinfo {author} {\bibnamefont {{Wonjoo Suh}}},
      \bibinfo {author} {\bibnamefont {{Zheng Wang}}}, \ and\ \bibinfo {author}
      {\bibnamefont {{Shanhui Fan}}},\ }\href {\doibase 10.1109/JQE.2004.834773}
      {\bibfield  {journal} {\bibinfo  {journal} {IEEE Journal of Quantum
      Electronics}\ }\textbf {\bibinfo {volume} {40}},\ \bibinfo {pages} {1511}
      (\bibinfo {year} {2004})}\BibitemShut {NoStop}%
    \bibitem [{\citenamefont {Haus}(1984)}]{Haus1984}%
      \BibitemOpen
      \bibfield  {author} {\bibinfo {author} {\bibfnamefont {H.~A.}\ \bibnamefont
      {Haus}},\ }\href@noop {} {\emph {\bibinfo {title} {{Waves and Fields in
      Optoelectronics}}}}\ (\bibinfo  {publisher} {Prentice Hall},\ \bibinfo {year}
      {1984})\ pp.\ \bibinfo {pages} {1--402}\BibitemShut {NoStop}%
    \end{thebibliography}

%

\end{document}